\documentclass{aa}

\usepackage{multirow}
\usepackage{comment}
\usepackage{graphicx}
\usepackage{circledsteps}
\usepackage{longtable}
\usepackage{txfonts}
\usepackage[]{hyperref}

\begin{document} 

\title{Detection of period-spacing patterns due to the gravity modes of rotating dwarfs in the TESS southern continuous~viewing~zone\thanks{The catalog file of stars with period-spacing patterns is available in electronic form at the CDS via anonymous ftp to \url{cdsarc.u-strasbg.fr} (130.79.128.5) or via \url{http://cdsarc.u-strasbg.fr/viz-bin/cat/J/A+A/662/A82}. The supplementary material, that presents the period-spacing pattern in our catalog, is available at \url{https://doi.org/10.5281/zenodo.6320990}}}

\author{
    S.~Garcia\inst{\ref{ivs}}
    \and
    T.~Van Reeth\inst{\ref{ivs}}        
    \and
    J.~De Ridder\inst{\ref{ivs}}
    \and
    A.~Tkachenko\inst{\ref{ivs}}
    \and
    L.~IJspeert\inst{\ref{ivs}}
    \and
    C.~Aerts\inst{\ref{ivs},\ref{Nijmegen},\ref{mpia}}
}

\institute{
    Instituut voor Sterrenkunde (IvS), KU Leuven, Celestijnenlaan 200D, B-3001 Leuven, Belgium \\
    \email{stefanorenzo.garciacastaneda@kuleuven.be}
    \label{ivs}
    \and
    Department of Astrophysics, IMAPP, Radboud University Nijmegen, PO Box 9010, 6500 GL, Nijmegen, The Netherlands
    \label{Nijmegen}
    \and 
    Max Planck Institute for Astronomy, Koenigstuhl 17, 69117, Heidelberg, Germany
    \label{mpia}
}

\date{Accepted November 22, 2021}

  \abstract
   {The theory of stellar evolution presents shortcomings when confronted with  asteroseismic probes of interior physical properties. The differences between observations and theory are often great because stellar models have mainly been calibrated from observables connected to the surface of stars. Period-spacing patterns caused by gravity modes are a particularly powerful asteroseismic tool that are useful for probing the near-core rotation and mixing of chemical elements in main-sequence stars with convective cores.}
   {We aim to compose a catalog of intermediate-mass stars 
   in the Transiting Exoplanet Survey Satellite (TESS) southern continuous viewing zone (CVZ) to reveal period-spacing patterns caused by gravity modes for use in future asteroseismic modeling.}
   {TESS full frame images (FFI) were inspected to select stars of intermediate- and high-mass using color-magnitude criteria. Light curves were extracted from custom masks per star, adopting stringent constraints on the aperture masks and contamination. The extracted light curves were subject to iterative prewhitening to detect gravity modes. We developed a method relying on the assumption that period spacings are an approximately linear function of the mode periods to build a template pattern. This template was used to extract the patterns and their uncertainties, relying on a bootstrap approach.}
   {Our TESS catalog of high-quality period-spacing patterns is the first of its kind and contains 140 gravity-mode patterns in 106 $\gamma\,$Dor stars and two slowly pulsating B-type (SPB) stars. Half of these patterns contain seven or more measured mode periods and the longest pattern contains 20 modes. We provide the community with a convenient software tool to search for period-spacing patterns and to process the extracted light curves.}
   {Our catalog offers a fruitful starting point for future gravity-mode asteroseismology of rotating dwarfs with convective cores in the southern hemisphere.}


\titlerunning{Period-spacing patterns in the TESS southern CVZ}
\authorrunning{Garcia et al.}

\maketitle

\section{Introduction}

The theory of stellar structure and evolution is well established and capable of describing the different stages of a star throughout its life in general terms \citep[e.g.,][]{Kippenhahn2012}. However, the theory is mostly calibrated to the surface properties of stars, such as their surface gravities, surface chemical compositions, surface rotations, and effective temperatures. Today, advances in asteroseismology and the advent of high-precision space photometry from telescopes such as CoRoT \citep{Auvergne2009},
{\em Kepler} \citep{Koch2010} and TESS \citep{Ricker2015} allow us to probe stellar interiors with a precision that cannot be reached from extrapolating surface quantities \citep[e.g.,][for recent reviews]{Hekker2017,GarciaBallot2019,Bowman2020c,Aerts2021}. Asteroseismic modeling based on space photometry has revealed large discrepancies between observations and the theory of stellar structure and evolution, such as in the transport of angular momentum \citep[e.g.,][]{Aerts2019}.

Gravity modes (hereafter g~modes) are stellar oscillations that have buoyancy as their dominant restoring force and have optimal probing power in the near-core regions of main-sequence stars \citep[e.g.,][]{Aerts2018}. They are detected in main-sequence stars with a convective core and a radiative envelope known as $\gamma$~Doradus ($\gamma$\,Dor) stars \citep{Kaye1999} and SPB stars \citep{Waelkens1991}, which have masses from 1.3 to 1.9\,$M_\odot$ and 3 to 9\,$M_\odot$, respectively \citep[e.g.,][Chapter\,2, for more properties]{Aerts2010}. To detect and accurately measure the frequencies of their individual g~modes, which have periodicities of the order of days, high-precision long-term uninterrupted observations are needed. These requirements are met for time-series photometry from space missions such as {\em Kepler} and TESS, which allows us to detect g-mode period spacing patterns. The first such detection in a dwarf was made for the SPB star HD\,50230 from CoRoT space photometry by \citet{Degroote2010}. The five-month CoRoT light curve was sufficient to detect a period-spacing pattern of eight dipole g~modes thanks to this star's very slow rotation, also justifying  asteroseismic modeling having ignored the Coriolis acceleration in the pulsation computations \citep{WuLi2019}. 

Gravity-mode period spacing patterns, describing the difference in periods of g~modes with an identical spherical degree, $\ell,$ and azimuthal order, $m,$ while having a consecutive radial order, $n$, are a powerful asteroseismic tool \citep[][for a detailed theoretical derivation]{Aerts2010}. This tool allows us to probe the near-core regions of main-sequence stars with convective core and a radiative envelope. As shown by \citet{Shibahashi1979} and \citet{Tassoul1980}, g-mode periods $P_{n\ell}$ are equidistant for a spherical chemically homogeneous non-rotating star when considering the asymptotic regime, that is with $2\pi/P_{n\ell} \ll N$ with $N$ the buoyancy frequency:
\begin{equation} \label{Eq:asymptomatic_approx}
P_{nl} = \frac{\Pi_0}{\sqrt{\ell(\ell+1)}}\, (n+\epsilon_g),
\end{equation}
with 
\begin{equation} \label{Eq:buoyancy_radius}
    \Pi_0 = \frac{2\pi^2}{\int_{r_1}^{r_2}N(r)\,r^{-1}\,\mathrm{d}r},
\end{equation}
where $\epsilon_g$ is a phase term which depends on the boundaries, $r_1$ and $r_2$, of the g-mode propagation cavity. Gradients in the stellar chemical composition profile cause mode trapping, which introduces wave-like behavior and periodic dips also known as buoyancy glitches in the period-spacing patterns. As shown by \citet{Miglio2008}, the amplitude and location of these modulations in the pattern depend on the steepness and the location of the chemical gradients inside the star.

For a rotating star, the Coriolis acceleration and the difference between the corotating reference frames and the inertial ones of the observer shift the g-mode frequencies quite drastically. As a result, the observed period spacings reveal a decreasing trend for prograde ($m > 0$) and zonal ($m = 0$) modes when plotted as a function of increasing mode period as observed in an inertial frame of reference. For retrograde modes ($m < 0$), on the other hand, an overall increase in the observed spacings of modes with increasing pulsation period occurs \citep[e.g.,][]{Bouabid2013,VanReeth2015a,VanReeth2015b,Ouazzani2017}. 
\citet{SchmidAerts2016} investigated the limit of the rotation frequency at which the Coriolis acceleration can still be treated perturbatively and found this approximation already breaks down for rotation frequencies above roughly $\sim\!0.1\,$d$^{-1}$, which is the case for almost all intermediate-mass dwarfs \citep[cf.\,][]{Aerts2017}.

A strong magnetic field in the stellar interior requires the Lorentz force to be included in the pulsation computations. This modifies the morphology of the observed period spacing pattern by introducing a saw-tooth modulation of the period spacing values as the consecutive pulsation periods increase \citep{Prat2019a,Prat2020a,JordanVanBeeck2020}. Moreover, coupling between inertial modes in the rotating core and g~modes in the envelope occurs and may cause dips superposed to buoyancy glitches at particular mode periods in the spacing diagram \citep{Ouazzani2020,Saio2021,Lee2021}. This implies that interpretations of buoyancy glitches from mathematical analytical descriptions ignoring the Coriolis acceleration \citep[e.g.,][]{Cunha2019} can only be applied to slowly-rotating non-magnetic stars. The majority of g-mode pulsators require the modeling and interpretation of the observed period spacing patterns to be done numerically, based on the inclusion of the Coriolis (and perhaps Lorentz) force when solving the pulsation equations \citep[cf.\,][]{Townsend2013,Townsend2018}. 

After the initial discovery by \citet{Degroote2010}, it took another five years before period-spacing patterns were detected from four-year {\it Kepler\/} light curves in several hundreds of $\gamma$\,Dor stars \citep[e.g.,][]{VanReeth2015b,LiGang2019a,GangLi2020} and several dozens of SPB stars \citep[e.g.,][]{Papics2017,Pedersen2021,Szewczuk2021}. These patterns have been used to measure the near-core rotation rates of all these stars \citep[e.g.,][]{VanReeth2016,Christophe2018,VanReeth2018,LiGang2019a,GangLi2020,Takata2020,Pedersen2021} and place constraints on the chemical transport processes that take place in the deep stellar interior \citep[e.g.,][]{Mombarg2020,Mombarg2021,Pedersen2021}.
From the point of view of improving stellar evolution theory, dwarfs are the most interesting targets as they still have their evolved stages ahead of them and uncertainties in the transport processes are cumulative over time. Moreover, g~modes are potentially excited along the main sequence in various instability regions for stars born with a convective core covering a broad mass range \citep[][Chapter\,3]{Aerts2010}. This is why we focus our work on dwarfs covering spectral types between O and F.

To date, the photometric observations obtained with the nominal {\em Kepler} mission cover the longest time base and are more precise than the observations from any other high-cadence space-photometry mission. Hence, most breakthroughs in g-mode asteroseismology of dwarfs were achieved thanks to {\em Kepler} observations so far. Here, we exploit the asteroseismic potential of the ongoing TESS space mission and compare it to  that of {\em Kepler}. The TESS extended mission is gradually providing data of progressively higher frequency resolution and has opened the door to analyzing numerous stars located in regions of the sky other than the {\em Kepler} field of view and in different metallicity regimes. In this work, we present results based on the first full year of uninterrupted TESS monitoring, to evaluate its capacity for the g-mode asteroseismology of rotating dwarfs. Future works will involve the addition of data from the extended TESS mission to the stars in our current catalog to increase their capacity for asteroseismic modeling.

Our study is aimed at identifying new $\gamma$\,Dor or SPB stars that have been observed by TESS in the southern continuous viewing zone (CVZ) to build the first TESS catalog of high-quality g-mode period-spacing patterns for such pulsators.
The southern CVZ was observed uninterruptedly during the first year of the nominal TESS mission, with a $24^\circ$ square field-of-view centered at the Southern ecliptic pole rotating about every 27\,d. This long observation period has provided light curves with a nominal frequency resolution of about $0.003\,\mathrm{d}^{-1}$.

The paper is organized as follows. In Section\,\ref{Sec:Data_set}, we describe our criteria for selecting O/B- and A/F-type stars in the TESS southern CVZ. In Section\,\ref{Sec:pipeline}, we describe our method used to extract light curves from the TESS full frame images, including our data analysis treatments to detrend and optimize the extracted light curves to search for g~modes. In Section\,\ref{Sec:freq_analysis}, we discuss the frequency extraction from the light curves and our posterior analysis. Section\,\ref{Sec:PSP-seach} describes our method for finding period-spacing patterns. Finally, we discuss our final catalog of g-mode pulsators with period-spacing patterns in Section\,\ref{Sec:PSP-catalog}. 
\section{Data set} \label{Sec:Data_set}

To select our sample of stars, we started from the TESS Input Catalog (TIC) version 8 \citep{Stassun2019} and reduced it to the TESS southern CVZ by imposing an ecliptic latitude $\beta \le -72^\circ$. To exclude extended objects and keep only point-like sources, we used the TIC's flag \texttt{Objtype=star}. Stars likely to be white dwarfs or giants were identified with the TIC's flag \texttt{wdflag=1} and \texttt{lumclass=GIANT}, respectively, and excluded from the sample. The first flag represents a cut in absolute {\em Gaia} magnitude and {\em Gaia} color ($G_{BP}-G_{RP}$) while the second flag represents a cut in radius (calculated from {\em Gaia} parallaxes) and $T_{\text{eff}}$. We refer to \citet{Stassun2019} for a description of the TIC flags.

\begin{figure}[h]
\centering
\includegraphics[width=0.49\textwidth]{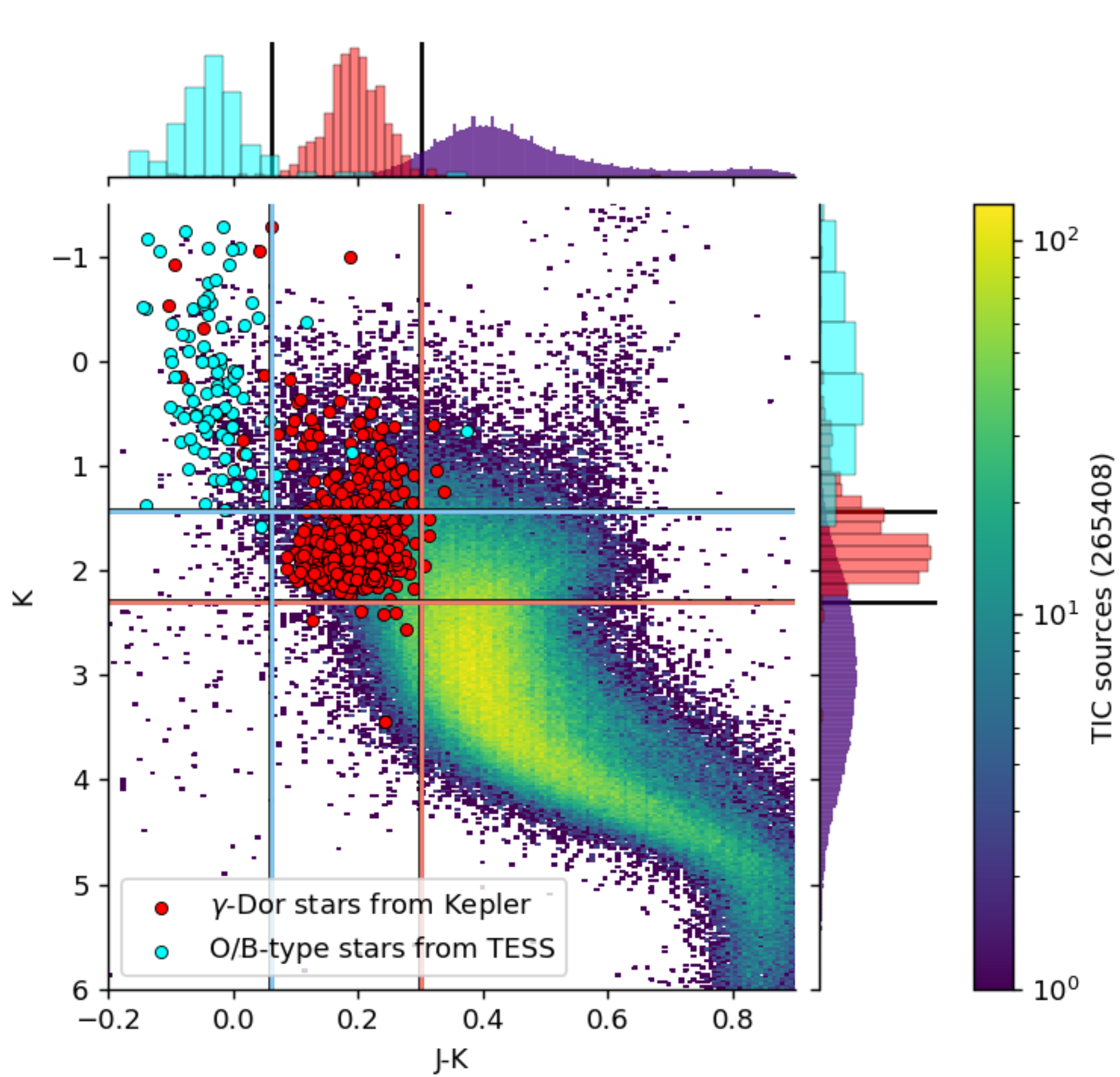}
\caption{2MASS color-magnitude diagram showing the pairs of cuts in absolute magnitude and color defining candidate g-mode pulsators of spectral type O/B (cyan straight lines) and A/F (red straight lines) in $K$ and $J-K$. The cyan and red circles are bona fide O/B- and A/F-type stars, respectively, while the retained dwarfs from the TIC in the southern CVZ are plotted in the background. The side histograms show the distribution of stars; the pairs of cuts enclose 95\% of the respective bona fide stars.}
\label{Fig:cutsKJK}
\end{figure}

To narrow down our sample of stars to candidates of spectral type F and hotter, that is the most likely g-mode pulsators, we used 139 TESS O/B-type stars selected manually by   \citet{Pedersen2019} and 616 {\em Kepler} A/F-type $\gamma\,$Dor pulsators taken from \citet{Tkachenko2013,GangLi2020}. We placed those stars in a color-magnitude diagram and used them to define two pairs of color-magnitude cuts that enclose 95\% of these bona fide O/B- and A/F-type stars. By applying these pairs of cuts to our TESS sample, we extracted all the O/B- and A/F-type candidate pulsators of interest. We calculated their absolute magnitudes as 
\begin{equation}
    M=m-5\log(d)+5,
\end{equation}
where $m$ is the apparent magnitude in a given passband. To obtain the distance $d$, we used the Bayesian distance estimate from \citet{Bailer-Jones2018} reported in the TIC. To ensure reliable distances, we used stars with a positive {\em Gaia} parallax of relative error of less than 25\% that passed both astrometric and photometric criteria given by Eqs.\,(1) and (2) in \citet{Arenou2018}, namely, with TIC flag \texttt{gaiaqflag=1}. To minimize the effect of extinction, we used the 2MASS infrared bands $J$, $H,$ and $K$ and adopted the cuts listed in Table\,\ref{Tab:CMD_Cuts}. Figure\,\ref{Fig:cutsKJK} shows these cuts in $K$ and $J-K$  as straight lines and our sample in the background. A/F-type candidates correspond to stars in the top-left quadrant delineated by the red straight lines minus the O/B-type candidates that correspond to stars in the top-left quadrant delineated by the cyan straight lines. The final candidates are obtained after an analogous additional selection in a color-magnitude diagram based on $H$ and $J-H$. Table\,\ref{Tab:CMD_Cuts} and Figure\,\ref{Fig:cutsKJK} do not consider corrections for extinction. The potential contamination by cooler stars that are not expected to pulsate in g~modes will be treated in Section\,\ref{Sec:freq_analysis}, based on the frequency analysis results.

\begin{table}
\caption[]{\label{Tab:CMD_Cuts}Cuts used to define O/B- and A/F-type candidates.}
\begin{center}
\begin{tabular}{ c|cc|cc } 
\hline
\hline
Spectral type candidate &  $K$ & $J-K$ & $H$ & $J-H$ \\
\hline
O/B & 1.429 & 0.06 & 1.55 & 0.045 \\
A/F & 2.300 & 0.30 & 2.35 & 0.240 \\
\hline
\end{tabular}
\tablefoot{Pairs of cuts in absolute magnitude and color in the 2MASS system (uncorrected for extinction) used to define O/B- and A/F-type candidates. The pairs of $K$ and $J-K$ are displayed in Figure\, \ref{Fig:cutsKJK}.}
\end{center}
\end{table}

Finally, to favor a high signal-to-noise ratio (S/N) and non-contaminated flux in the light curves, we limited our sample further to stars with apparent TESS magnitude brighter than 15 (uncorrected for extinction) and situated at least 2 arcsec apart from other stars in the TIC. Our selected sample consists of 345 O/B-type candidates and 9369 A/F-type candidates in the TESS southern CVZ, all from our Galaxy.

\section{Our TESS data reduction pipeline}\label{Sec:pipeline}

\begin{figure*}
\resizebox{\hsize}{!}
{\includegraphics[width=\hsize,clip]{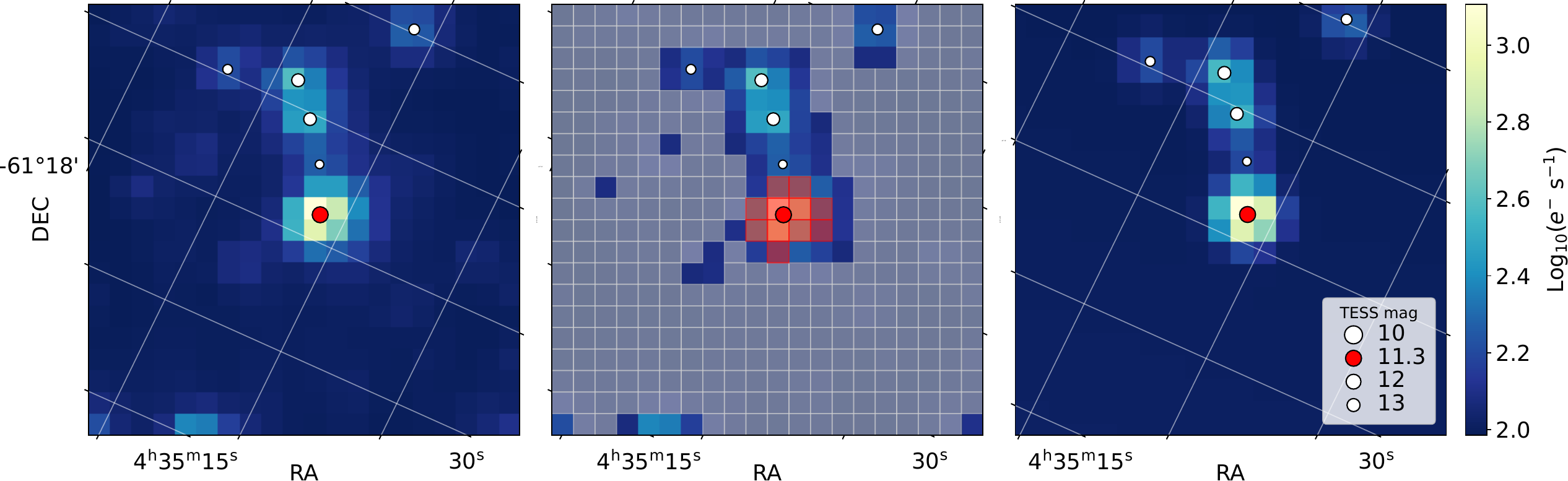}}
\caption{Square 20-by-20 pixels around TIC 38845463, TESS sector 1. The color bar is common to the three panels and shows a logarithmic scale of the flux in electrons per second. The red circle represents the target star while white circles indicate TIC neighboring stars down to four TESS magnitudes fainter with respect to the target star. The symbol sizes are  inversely proportional to the TESS magnitude. Declination grids are $2'$ apart. {\em Left:}  Median image of the target pixel file. {\em Middle:} Panel with the final aperture mask (red shade) and background mask (gray shade) overplotted. The aperture mask results from the threshold parameter $n=20$ (see text for explanation). {\em Right:} Best fit based on the left panel used to estimate the level of contamination in the aperture mask due to flux from neighboring stars. The image was modeled as six 2-D Gaussian functions plus a 2-D plane. See Section\,\ref{Sec:masks} for further details.}
\label{Fig:tpfs}
\end{figure*}

\begin{figure*}
\resizebox{\hsize}{!}
{\includegraphics[width=\hsize,clip]{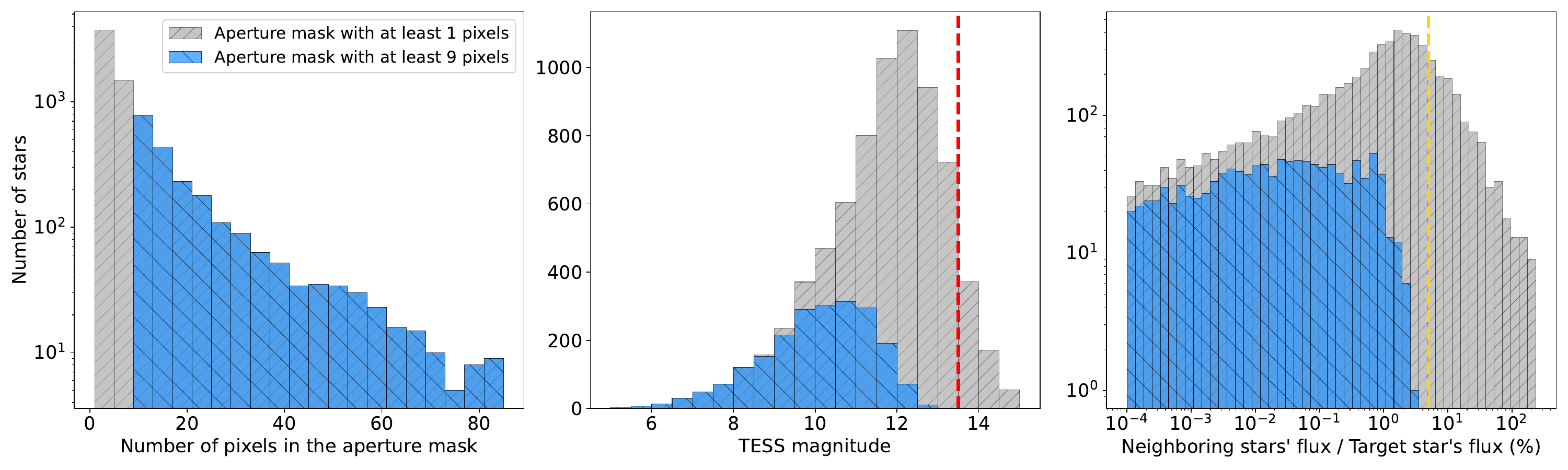}}
\caption{Our star sample without constraints on the aperture mask size (7385 sources, gray histogram) and after imposing a minimum size of 9 pixels (2162 sources, blue histogram). Other constraints described in Section\,\ref{Sec:masks} apply to both histograms. {\em Left:}  Sample distribution of the median aperture mask sizes, calculated per star over all TESS sectors. {\em Middle:}  Sample distribution of the TESS magnitudes. The dashed red line marks a magnitude value of 13.5. {\em Right:} Median contamination level caused by neighboring stars, calculated per star over all TESS sectors. The dashed orange line marks a contamination of 5\%.}
\label{Fig:sample_histograms}
\end{figure*}

We searched sectors 1 to 13 of TESS for long-cadence (i.e., 27.4 minutes) full-frame images available in the Mikulski Archive for Space Telescopes and used the \texttt{TESScut} API \citep{Brasseur2019} to download, for every star in our sample, a 20X20 pixel image with the target star at the center. These images are known as target pixel files and contain the flux information for all available time stamps. The 20X20 pixel size was chosen such that the target pixel file contains both the flux of the target star and the flux of the representative background around it; a typical example is shown in Figure\,\ref{Fig:tpfs} in which the middle panel shows a background mask 11 times larger than the aperture mask. Light curves were extracted from the target pixel files using aperture photometry, as explained in Section \ref{Sec:masks}, while the background and systematic effects were corrected using two standard statistical methods, that is, 
a principal component analysis (PCA) and a linear regression, as further detailed in Section \ref{subsec:lc}. The Python package Lightkurve \citep{lightkurve} was used during the reduction.

\subsection{Aperture and background masks}\label{Sec:masks}

To define the aperture mask of a star, we used the median frame of all the target pixel files (left panel in Figure\,\ref{Fig:tpfs}) and selected all pixels with a flux count larger than the median flux plus $n$-times the standard deviation. Those pixels are our first estimate of the aperture mask. The standard deviation was robustly estimated as 1.4826 times the median absolute deviation \citep{Ruppert2011}. To reduce the contamination from nearby stars falling into the aperture mask, we used the increasing values of $n=5.0, 7.5, 10, 15, 20, 30,$ and $40$ for the threshold in standard deviation to shrink the aperture mask until the target star was the only bright star contained within it by at least 4 TESS magnitudes. The target and neighboring stars with apparent TESS magnitudes $m_{\textrm{TESS}}^{\textrm{target}}$ and $m_{\textrm{TESS}}$, respectively, within the aperture mask thus follow the condition:
\begin{equation}\label{Eq:mask}
m_{\textrm{TESS}} - m_{\textrm{TESS}}^{\textrm{target}} \ge 4,
\end{equation}
ensuring that the flux of individual fainter stars contributes no more than approximately $0.25\%$ of the total flux within the aperture mask. For cases where the resulting aperture mask consists of disjointed sets of pixels, only the set containing the target star is kept. Finally, to help prevent flux from neighboring stars leaking into the aperture mask, pixels showing an increase in flux in any direction away from the target star are removed from the mask and used as its edge. The background mask was defined in the same way as the first estimate of the aperture mask but selecting the pixels below a threshold with $n=3$, thus ensuring a minimum two-standard-deviation flux gap between the aperture mask and the background mask. A typical example of both final apertures is shown in the middle panel of Figure\,\ref{Fig:tpfs}.

To estimate the level of contamination in the aperture mask due to the flux of neighboring stars, we calculate the ratio of this flux to that of the target star. To obtain such fluxes, all stars complying with Eq. (\ref{Eq:mask}) were modeled as 2D Gaussian functions and fitted to the median image of the target pixel file along with a 2D plane to account for the background flux. The Gaussian functions were centered at the location of each star, had all the same standard deviation, and their relative amplitudes kept the same relation as the fluxes from the stars. Fluxes were converted from TESS magnitudes using the TASOC Photometry pipeline \citep{Handberg2021}. The right panel in Figure\,\ref{Fig:tpfs} shows an example of this fit.

We rejected the aperture mask (together with the target pixel file) when the final mask contained stars that do not comply with Eq. (\ref{Eq:mask}). To avoid both corrupted and bleeding pixels, we also rejected masks that had pixels with null counts or that were too elongated (i.e., with fewer than four rows or columns, while the other is at least 14-pixels long). To average out the stochastic noise of individual pixels, we only kept aperture masks with at least nine pixels, as shown in the left panel of Figure\,\ref{Fig:sample_histograms}. After careful assessment, our sample consisted of 2162 stars with a flux contamination due to neighboring stars smaller than 5\%, as shown in the right panel in Figure\,\ref{Fig:sample_histograms}. The middle panel in Figure\,\ref{Fig:sample_histograms} shows that our light curve extraction is consistent with previous data pipelines for which the extraction is considered to be trustworthy only for stars brighter than TESS magnitude of 13.5 \citep[e.g.,][]{Handberg2019,Huang2020,Caldwell2020}.

\begin{figure}[h!]
\centering
\includegraphics[width=\hsize,clip]{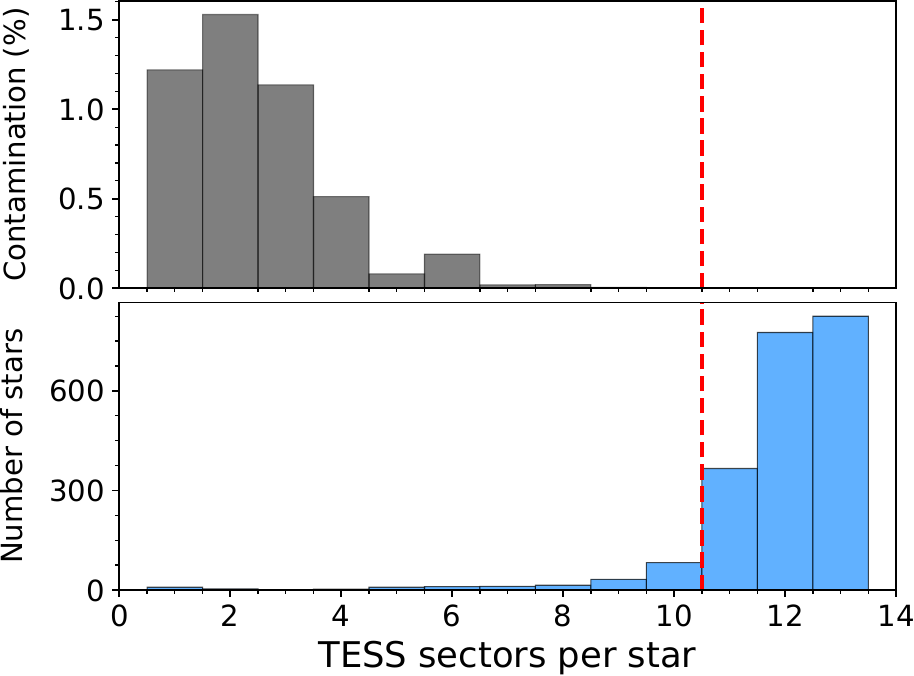}
\caption{Blue histogram in the bottom shows the number of TESS sectors per star with a satisfactory aperture mask. The dark-gray histogram shows the mean contamination in each bin of the blue histogram. The dashed red line shows the cut for stars with at least 11 TESS sectors (a total of 1967 stars).}
\label{Fig:hist_tess_sectors}
\end{figure}

Because of our stringent constraints on the aperture mask, not all TESS sectors yielded a satisfactory mask for a given star. We therefore only kept stars with aperture masks found in at least 11 of the 13 TESS sectors,
as shown by the dashed line in  Figure\,\ref{Fig:hist_tess_sectors}. The 11 sectors are not necessarily consecutive. After this cut, our sample consists of 1967 stars. Figure\,\ref{Fig:hist_tess_sectors} also shows a higher level of contamination when the aperture mask is found in fewer numbers of TESS sectors, indicating that stars in a crowded field are more prone to fail the requirements of the aperture mask selection.

\subsection{Light curve extraction and correction} \label{subsec:lc}

\begin{figure}[h!]
\centering
\includegraphics[width=0.48\textwidth]{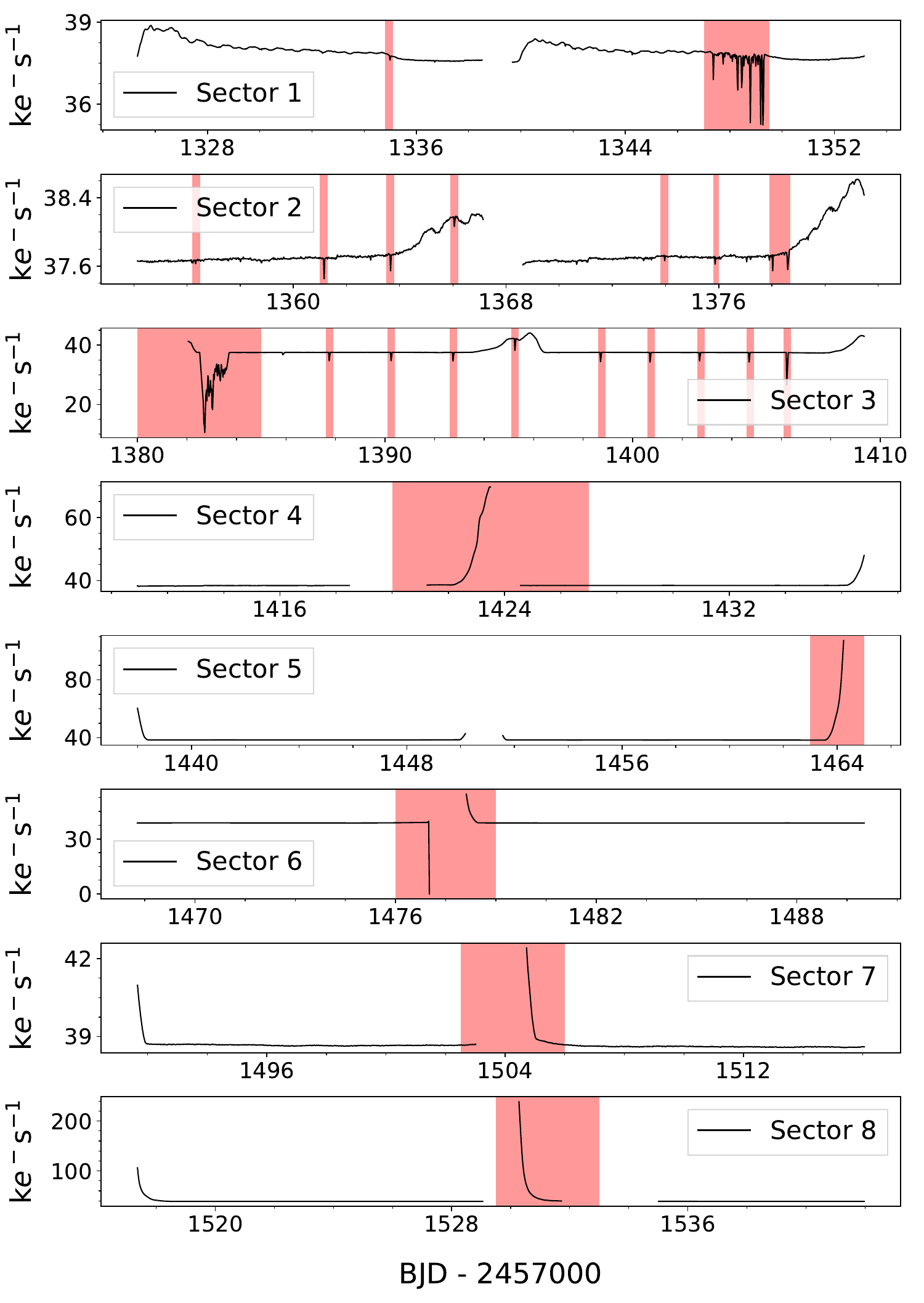}
\caption{Uncorrected light curves of TIC 30192406, showing in red the time intervals that have been excluded for all stars according to Table\,\ref{Tab:excluded_intervals}. Sector 1 shows an example of jittering of the satellite. Sector 2 shows an example of scattered sunlight reflected by the Earth or the Moon. Sector 3 shows an example of systematic flux variability caused by the periodic re-pointing of the camera.}
\label{Fig:excluded_intervals}
\end{figure}

To remove part of the systematic flux variability in the extracted light curves, we used the data quality flags provided in the headers of the target pixel file\footnote{Descriptions about TESS quality flags can be found in the section ``Data Product Overview'' in \url{https://outerspace.stsci.edu/display/TESS}.} \citep{Twicken2020} and removed the data from the time stamps affected by loss of fine pointing\footnote{Flagged as attitude tweak, coarse pointing and desaturation event.}, thermal transients\footnote{Flagged as safe mode.}, Earth pointing, and other specific effects\footnote{Flagged as manual exclude.} (e.g., coronal mass ejections). We then extracted the light curves using simple aperture photometry with the aperture masks we constructed as described in Section\,\ref{Sec:masks}. An example is shown in Figure\,\ref{Fig:excluded_intervals}, where the gaps in the data correspond to the removed time stamps. We noted that the use of the quality flags according to the TESS release notes did not cover all systemics present in the light curves and proceeded to manually remove time intervals (common to all stars) which still were significantly affected by systematic effects (e.g., telescope jittering, thermal transients, and loss of fine pointing). Such time intervals are present in sectors 1 to 8, as listed in Table\,\ref{Tab:excluded_intervals} and shown in red in Figure\,\ref{Fig:excluded_intervals}.

\begin{figure}[h!]
\centering
\includegraphics[width=0.93\hsize,height=5.6cm]{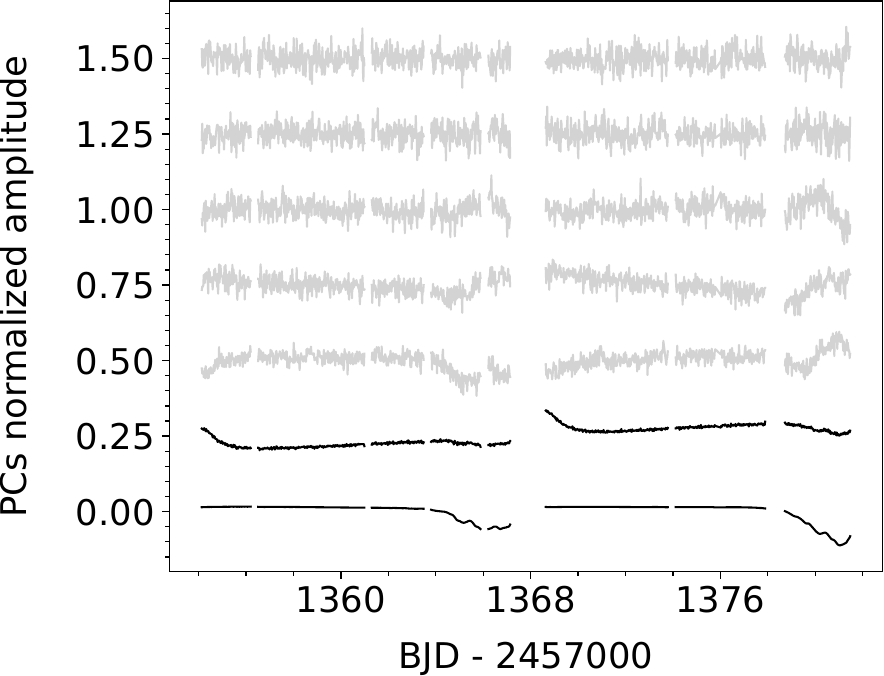}
\caption{First seven normalized principal components (PC; columns of the matrix $\mathbf{U}$) from the background mask of TIC 374944608, sector 2. The PCs are displayed in ascending order with the first normalized PC at the bottom and manually set 0.25 units apart from each other for a better visualization. Only black normalized PCs have a level of scattering $<10^{-4}$ (see Section \ref{subsec:lc}) and were used for the detrending of the light curve.}
\label{Fig:PCA}
\end{figure}

\begin{figure*}[h!]
\centering
\includegraphics[width=\hsize,clip]{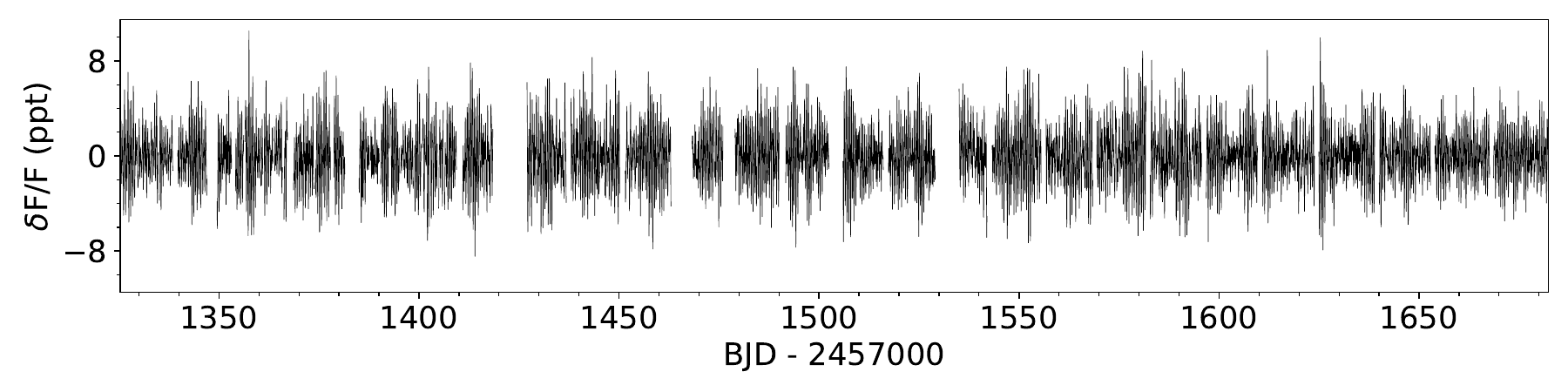}
\caption{Final light curve for TIC\,374944608 as derived from our developed pipeline discussed in Section\,3.}
\label{Fig:lc_example}
\end{figure*}

The remaining systematic variability (e.g., the gradual increase of flux in sector 2 due to scattered light or the rapid decrease of flux at the beginning of sector 8 as illustrated in Figure\,\ref{Fig:excluded_intervals}) and background flux were removed using a linear regression of the light curve against the flux variability present in the background mask defined in Section \ref{Sec:masks}. We started this process by extracting a light curve using aperture photometry from each of the background pixels (an average of 330 pixels per pixel target file in our sample). Subsequently, we applied a PCA to  these light curves to capture their dominant flux variability \citep[see][]{Feinstein2019}. We let $\mathbf{B}$ be the matrix whose columns are the light curves extracted from the background pixels, its singular value decomposition is
\begin{equation}
    \mathbf{B} = \mathbf{U}\textbf{S}\textbf{V}^T\;,
\end{equation}
where $\mathbf{U}$ and $\textbf{V}$ are orthonormal matrices and $\textbf{S}$ is a diagonal matrix with entries $s_i > s_{i+1}$. The principal components of $\mathbf{B}$ are given by the columns of $\mathbf{US}$, which are ordered by their contribution to the background flux variability. No universal number $k$ of principle components can be used to estimate background flux variability because different TESS sectors are subjected to different systematic effects. Since the first principal components capture most of the systematic variability while subsequent ones are affected by an increasing amount of white noise (see Figure \ref{Fig:PCA} for an example), we used the level of scatter in the normalized principal component (columns of $\mathbf{U}$) as a criterion to determine $k$. The level of scatter was determined by the median of the moving variances with a window length $W=16$ hours. Denoting the elements in a column of $\mathbf{U}$ by $f(t)$, where the time $t$ represents a cadence, the moving variance at a cadence $t$ was computed as
\begin{equation}
    \sigma^{2}_{L}(t) = \frac{1}{N-1} \sum^{N}_{|t_n - t| < W} \left(f(t_n) - \overline{f(t_n)}\right)^2\;,
\end{equation}
where $\overline{f(t_n)}$ is the mean of $f(t_n)$ within the window $|t_n - t| < W$. The $W=16$ hours was chosen to yield a large number of short light curve segments while being shorter than the typical period of g modes. We found that the level of scatter in the columns of $\mathbf{U}$, that is, $\vec{u_i}$, increased with $i$ as expected, since increasing values of $i$ had less systematic signal and more white noise. The level of scatter converged to a value of order $10^{-3}$ for $i \gtrsim 6$ regardless of the TESS sector. We therefore used the principal components with a level of scatter $<10^{-4}$ to represent the systematic variability present in the background. The median number of principal components used by our method is $k=4$. To  prevent further injection of white noise into the reduced light curves, we used a maximum of seven principal components.

Once the number $k$ of principal components was determined, we created the regressor matrix $\mathbf{X,}$ using the principal components as its columns and added a constant column to account for a constant level of background. Subsequently, we performed the following linear regression, assuming that the model fits the data within Gaussian uncertainties:
\begin{equation}
\vec{Y}=\mathbf{X}\vec{w}+\varepsilon,
\end{equation}
where $\vec{Y}$ represents the uncorrected light curve of the target star, $\vec{w}$ contains the regression coefficients, and $\varepsilon$ represents the noise. The corrected light curve was then computed as $\vec{Y}-\mathbf{X}\vec{w}$.
These corrected light curves were then normalized by subtracting their mean flux and then dividing them by it. Values further away than 5$\sigma$ were treated as outliers and removed (e.g., spikes due to cosmic rays). Finally, we stitched together the normalized light curves from each TESS sector as shown in Figure\,\ref{Fig:lc_example}.

\section{Frequency analysis from iterative prewhitening} \label{Sec:freq_analysis}

We analyzed the light curves resulting from our developed data analysis pipeline following a procedure of iterative prewhitening. \citet{VanBeeck2021} has offered an extensive description of five different prewhitening methods applied to g-mode pulsators, relying on various regression techniques and stopping criteria. Since \citet{VanBeeck2021} developed their methodology for g~modes in SPB stars and four-year {\it Kepler\/} light curves, their paper is highly relevant for our TESS work as well. We refer to that paper for a detailed description, as well as an elaborate comparative study of the efficiency of these five methods. Here, we rely on a method using the same frequency resolution restriction, a stopping criterion based on the amplitudes of the modes, and a nonlinear optimization to achieve the final regression result, as such methods were found to be the most powerful procedures based on the assumption of a harmonic fit to the light curve by \citet[][see methods 2 and 3]{VanBeeck2021}. We provide a summary of the adopted procedure, where frequencies were extracted in a 5-step process from the stitched light curves as the one shown in Figure\,\ref{Fig:lc_example}.

\begin{figure}[h!]
\centering
\includegraphics[width=\hsize,clip]{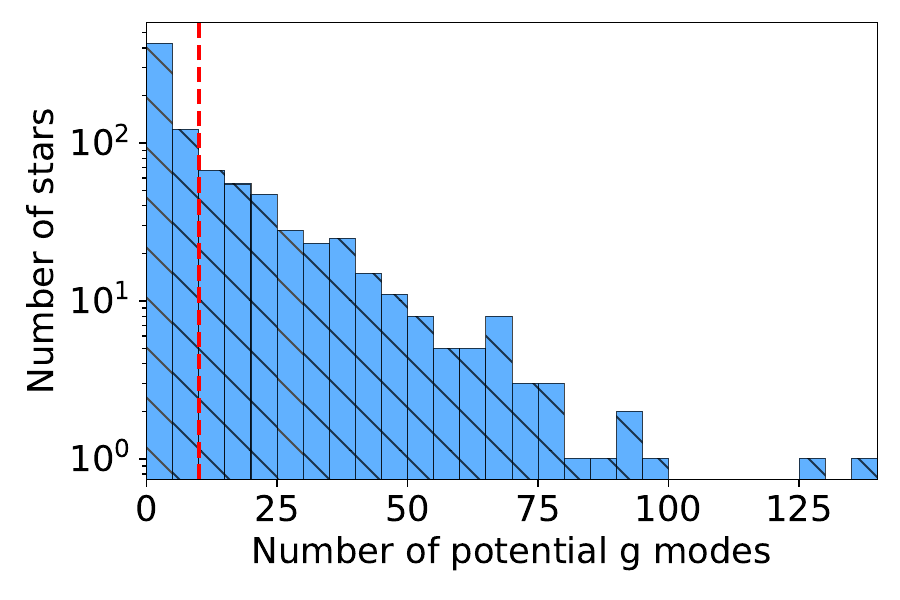}
\caption{Number of potential g-mode pulsations with $\mathrm{S/N}\ge4$ per star. Frequencies were deduced from Step 1 described in Section\,\ref{Sec:freq_analysis}. Our sample of 1967 stars has 1108 stars that are not in compliance with our Step 1 criterion and those are not plotted. The dashed red line shows the cut used to select rich enough period-spacing pattern candidates, namely, stars with at least ten significant mode periods. This resulted in 369 candidate  pulsators.}
\label{Fig:hist_gmodes}
\end{figure}

Step 1: We computed a Lomb-Scargle periodogram of the light curve using a 10-fold oversampled frequency range (compared to $T^{-1}$ with $T$ the time span of the light curve) from zero up to the Nyquist frequency. The frequency with highest amplitude is selected and a harmonic fit with that frequency to the light curve is determined, using linear regression. The best-fitting harmonic function is subtracted from the light curve and the process is iteratively repeated on the residual light curve until the selected frequency has an amplitude with $\mathrm{S/N}<4$. The noise level is calculated as the mean amplitude of the periodogram within a symmetric frequency window of size $1\ \mathrm{d}^{-1}$ centered on the selected frequency. An example of the periodogram for TIC 38845463 is visualized in panel \textbf{\Circled{A}} 
of Figure\,\ref{Fig:examples_of_results}
discussed further in the text.
In order to find candidates with g-mode period-spacing patterns, we only kept stars with at least ten significant potential g-mode frequencies, that is, those with pulsations of periods longer than 0.25 days. This restriction, illustrated in Figure\,\ref{Fig:hist_gmodes}, leaves us with a sample of 369 candidate stars.

Step 2: Unresolved frequencies were removed by requiring a conservative difference of at least $2.5 \times T^{-1}$ between extracted frequencies \citep{Loumos1978}. In case of unresolved frequencies, we kept the one with the largest amplitude.

Step 3: All accepted frequencies were optimized simultaneously using a Levenberg-Marquardt algorithm to perform a non-linear regression, with the output of the linear regression as initial input guesses. Uncertainties in the parameters were calculated following \citet{Montgomery1999} and corrected for their possible correlated nature following \citet{Schwarzenberg2003}. Frequencies whose corresponding amplitudes were consistent with zero within three standard deviations were considered as insignificant and rejected.

Step 4: To minimize the influence of the spectral window convolved with dominant frequencies in the periodogram during the iterative prewhitening process, we used the amplitude criterion developed by \citet{VanReeth2015a}:
\begin{equation} 
\alpha \leq \frac{A_{f}}{A_{l o c}} \leq \frac{1}{\alpha}\,,
\end{equation}
where $A_{f}$ and $A_{loc}$ are the optimized amplitudes and the amplitudes from the original Lomb-Scargle periodogram, respectively. This constraint is independent of the S/N and helps to avoid spurious frequency detections that can occur for space time-series data like {\em Kepler} and TESS with $\mathrm{S/N}>4$ \citep[e.g.,][]{Zong2016a,Baran2021,Bowman2021}. Moreover, \citet{VanBeeck2021} have shown criteria based on mode amplitudes to work better than just using S/N as a stop criterion. In practice, we used $\alpha=0.75,$ which was found to work optimally for TESS light curves by \citet{Antoci2019}.

Step 5: Combination frequencies were not considered as independent mode frequencies to hunt for period-spacing patterns. Such combination frequencies were identified through the following equation:
\begin{equation}  \label{Eq:combi_freq}
\left|f_k-\left( n_i f_i + n_j f_j \right) \right| \le \varepsilon\,,
\end{equation}
where we adopt the terminology of \citet{Degroote2009}, meaning that $f_i$ and $f_j$ are the parent frequencies, $n_i$ and $n_j$ are integer combination coefficients, $f_k$ is the combination frequency, and $\varepsilon$ is the threshold tolerance. We selected the 20 highest-amplitude frequencies per star as parent frequencies and searched for combinations of up to two parents, which leads to $|n_i|+|n_j| \le 2$.

\begin{figure}[h]
\centering
\includegraphics[width=\hsize]{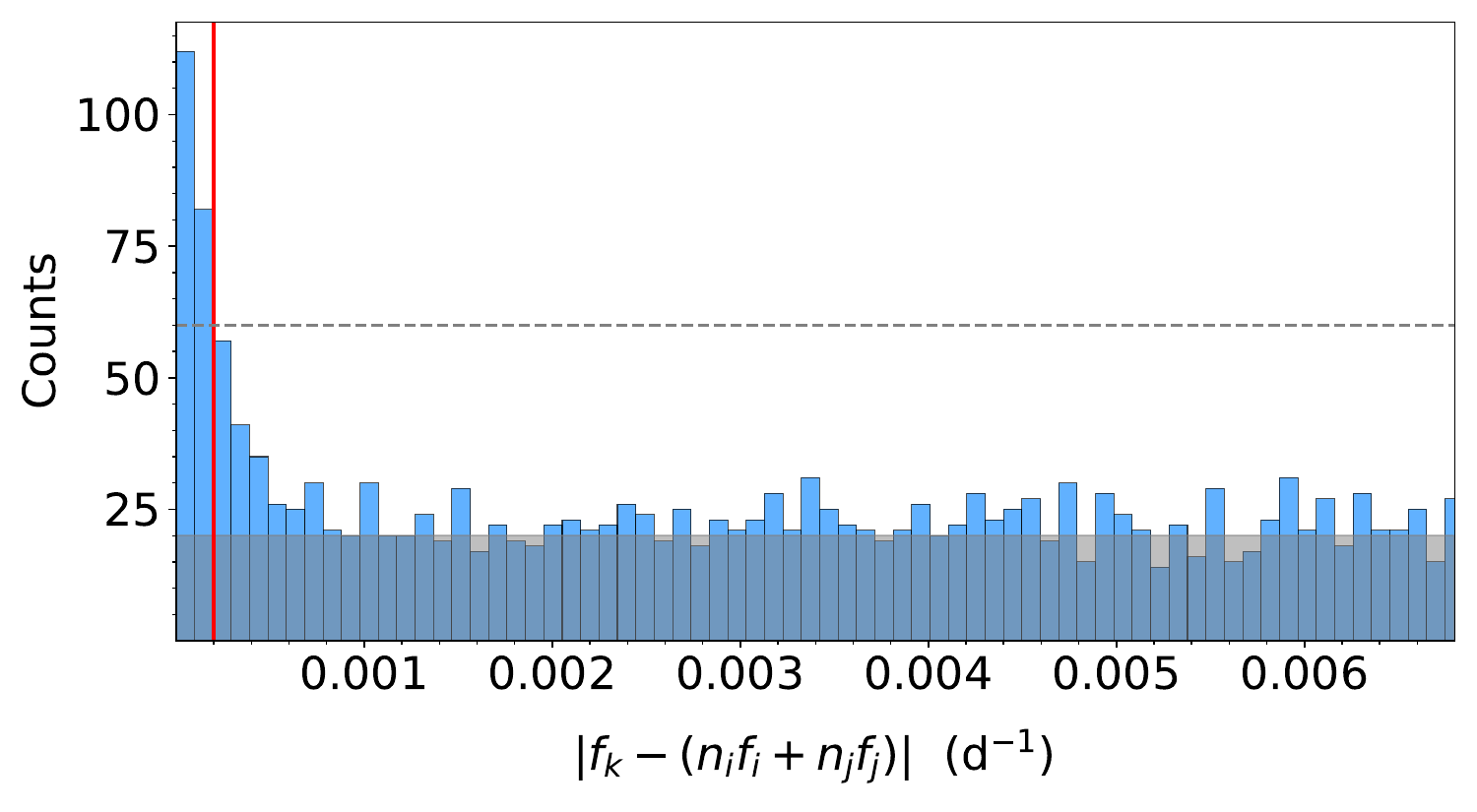}
\caption{Histogram following Eq.\,(\ref{Eq:combi_freq}) for the frequencies of stars in our g-mode sample. The gray shade marks the background level that represents a random match among combination frequencies and the horizontal dashed gray line marks 3 times that level. The vertical red shows the intercept of the dashed gray line and the distribution; it marks $\varepsilon=0.0002$ according to Eq.\,(\ref{Eq:combi_freq}). Frequencies occurring to the left of this line have a 67\% probability of corresponding to a genuine combination frequency.}
\label{Fig:combination_frequencies}
\end{figure}

Given the large number of frequencies per star (we note that Figure\,\ref{Fig:hist_gmodes} only counts g~modes and ignores p~modes), a linear combination of frequencies is likely to occur close to another independent frequency in the data without having to be a combination frequency \citep{Papics2012,Kurtz2015}. This is illustrated in Figure\,\ref{Fig:combination_frequencies} by the background count level marked with a gray shade while genuine combination frequencies correspond to the excess above such level. We therefore took three times that background level (gray dashed line in Figure\,\ref{Fig:combination_frequencies}) as a 67\% probability of being a genuine combination frequency. Such a probability corresponds to $\varepsilon=0.0002$ (red vertical line in Figure\,\ref{Fig:combination_frequencies}) and is consistent with the threshold tolerance reported by \citet{LiGang2019a}.

\begin{figure}[h]
\centering
\includegraphics[width=0.47\textwidth]{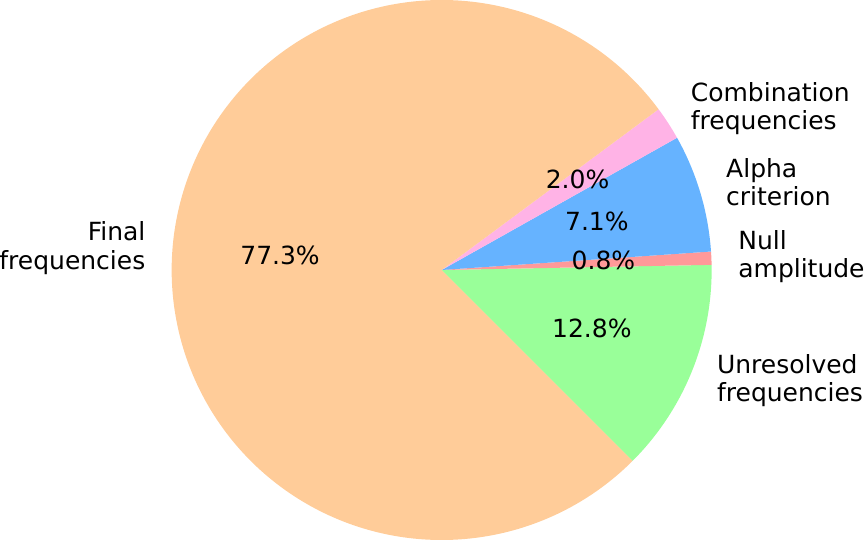}
\caption{Number of frequencies remaining in our sample after each step in Section\,\ref{Sec:freq_analysis} is applied. The pie chart is based on the frequencies of the 369 stars after step 1 of our frequency analysis. The total number of frequencies is 10927.}
\label{Fig:freq_analisys}
\end{figure}

\begin{figure*}[h!]
\centering
\includegraphics[width=\textwidth,clip]{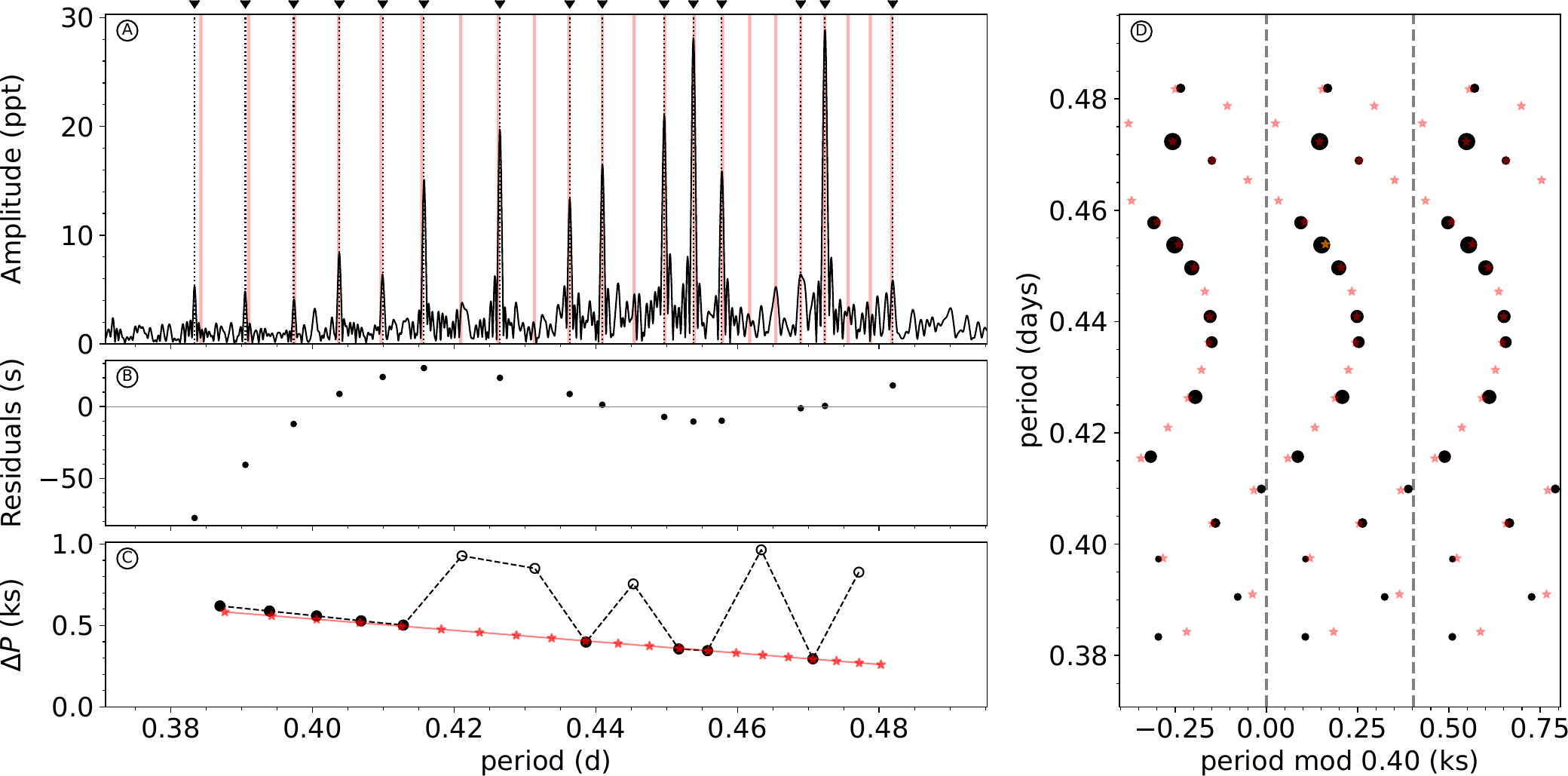}
\caption{Best-fit period-spacing pattern for TIC\,374944608. Plots generated with the interactive code \texttt{FLOSSY}. \textbf{\Circled{A} :}  Lomb-Scargle periodogram (solid black line), observed periods (vertical dotted black lines), best-fit linear template (vertical red lines). Observed periods used for the fit are indicated with black triangles at the top. \textbf{\Circled{B} :} Deviations from the linear pattern (i.e., difference between the periods indicated by the black triangles and the red vertical lines). \textbf{\Circled{C} :} Period spacing as a function of mode period. Both black and white circles are the observations. The red markers are the best-fit linear pattern with slope $\alpha$. Note that the fit is performed on $P$, not on $\Delta P$, and missing mode periods in the observations create artificially larger $\Delta P$ values (white circles). \textbf{\Circled{D} :} \'echelle diagram. The black circles are the periods used for the fit. The size of the circles is proportional to the amplitude of the amplitudes in the periodogram. The red markers are the best-fit linear pattern. The supplementary material contains a version of this figure for every pattern in our catalog.}
\label{Fig:examples_of_results}
\end{figure*}

Figure\,\ref{Fig:freq_analisys} shows the impact on the number of independent mode frequencies after each step of the frequency analysis. As an example, the final accepted frequencies of TIC\,374944608 (light curve in Figure\,\ref{Fig:lc_example}) are indicated with vertical dotted lines in 
Figure\,\ref{Fig:examples_of_results} (to be discussed below). We checked the 369 light curves of our sample after the frequency analysis and removed eclipsing binaries. These were studied further in the separate paper by \citet{IJspeert2021}. We also inspected the light curves for cases where systematic flux variability persisted after our data reduction pipeline (see Section \ref{Sec:pipeline}) and removed these from the sample. This inspection was done by eye, narrowing down our sample of period-spacing pattern candidates to 304 stars.

\section{Period-spacing pattern search}\label{Sec:PSP-seach}

\begin{figure*}[h!]
\centering
\includegraphics[width=\textwidth,clip]{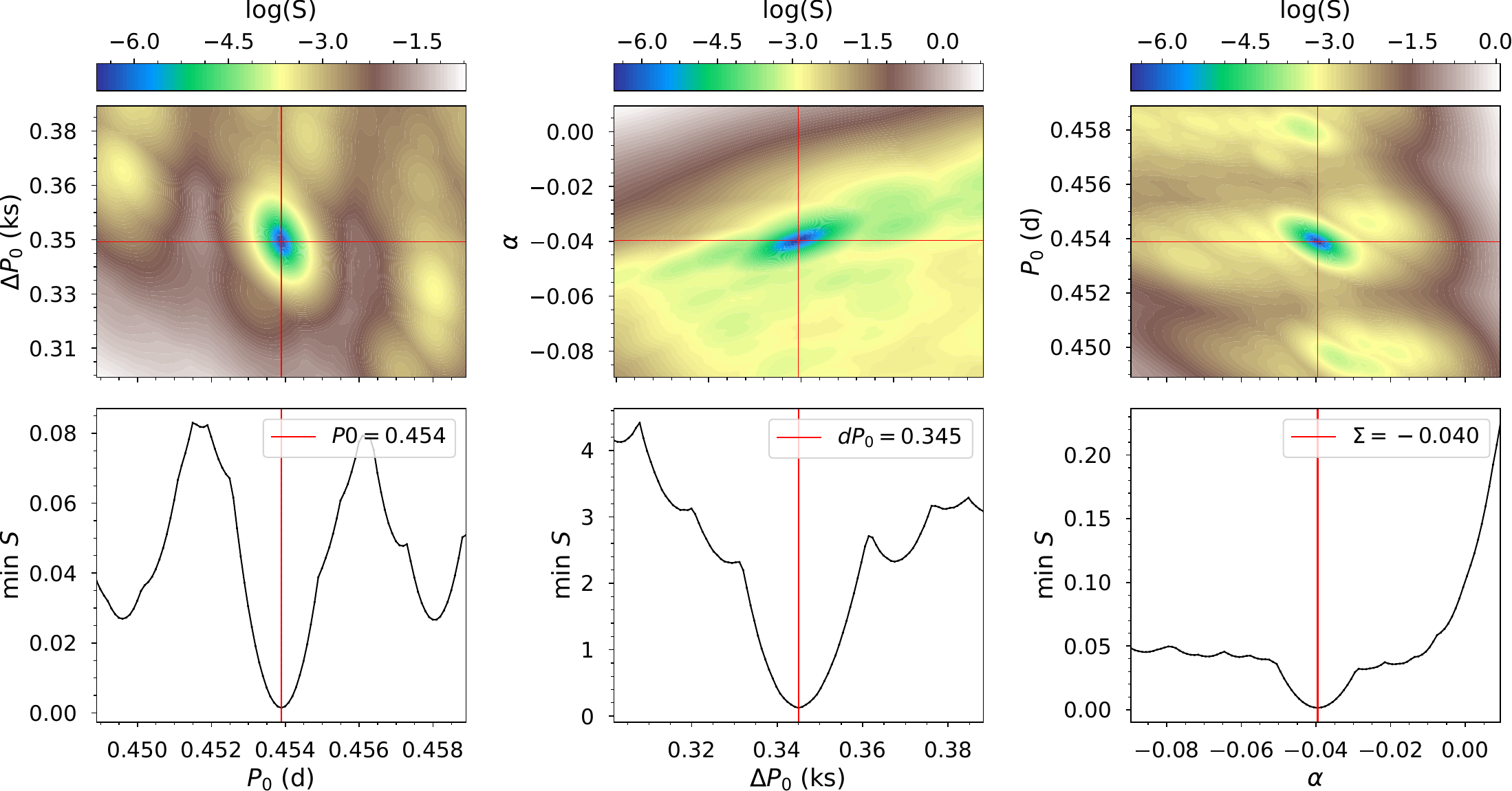}
\caption{Values of the cost function around the best-fit solution for TIC\,374944608 shown in Figure\,\ref{Fig:examples_of_results}. Plots generated with the interactive code \texttt{FLOSSY}. {\em Top:}  Correlation plots. {\em Bottom:} Minimum value of the cost function S as a function of the template period-spacing parameters. The supplementary material contains a version of this figure for every pattern in our catalog.}
\label{Fig:error_estimate}
\end{figure*}

To search for period-spacing patterns in our sample, we fitted the following template to the list of periods of each star \citep{LiGang2019a}:
\begin{equation} \label{Eq:pattern}
P_{i} = \sum_{j=0}^{i-1} \Delta P_j + P_0 = \Delta P_{0} \frac{(1+\alpha)^{i}-1}{\alpha}+P_{0}\;,
\end{equation} 
where $\Delta P_j = P_{j+1} - P_j$ is the local period spacing and $\Delta P_0$ is the period spacing at a reference period $P_0$. This period pattern allows for a linear change $\alpha\equiv\mathrm{d}(\Delta P)/\mathrm{d}P$ of the local period spacing caused by stellar rotation \citep{Ouazzani2017,Christophe2018a,LiGang2019a}.
The template depends on the three parameters $\{P_0, \Delta P_0, \alpha\}$. To account for the amplitude of the individual periods and the local size of the period spacing, we fitted Eq.\,(\ref{Eq:pattern}) to the observations, by minimizing the following custom cost function:
\begin{equation}  \label{Eq:cost_function}
S \left( P_0, \Delta P_0, \alpha \right)=\sum_{i=1}^{n} \frac{A_{i}}{A_{\rm max}} \frac{\left(P_{\,i}^{\mathrm{\;obs}}-{P}_{i}\right)^{2}}{\sigma_{i}^{2} + \Delta {P}_{i}^{2}}\;,
\end{equation}
where $P_i$ is the estimated period closest to the observed pulsation period $P^{\mathrm{\;obs}}_i$, $\Delta P_i$ is the estimated local period spacing, $A_i$ is the observed amplitude corresponding to $P_i^{\rm obs}$, $A_{\rm max}$ is the maximum observed amplitude, and $\sigma_i$ is the observed period uncertainty. Rather than minimizing the square of the absolute differences $(P_i^{\rm obs} - P_i)^2,$ we minimize the relative differences $(P_i^{\rm obs} - P_i)^2 / \Delta P^2_i$. In this way, period mismatches are more strongly penalized when they are large compared to the local period spacing. The addition of $\sigma_i^2$ in the denominator serves to limit the penalization when the local period spacing is comparable to the observational period uncertainty, $\sigma_i$. The extra weight, $A_{i}/A_{\rm max}$, serves to penalize a pattern more strongly when it mismatches the higher-amplitude mode periods.  The minimization of the cost function $S$ was done with the quasi-Newton method L-BFGS-B \citep{Byrd1995} implemented in the Python module \texttt{Scipy} \citep{SciPy}.

To find the location of the patterns in the periodogram as well as the initial guesses $\mathbf{\theta}^{\textrm{\;init}} = \{P_0^{\textrm{\;init}}, \Delta P_0^{\textrm{\;init}}, \alpha^{\textrm{\;init}} \}$ for the fit, we used two diagnostic plots to cover both the cases of rapid and slow rotators. Slow rotators show an approximately constant period spacing. Their period-spacing pattern can therefore be identified in an \'echelle diagram, where the period is plotted as a function of the period modulo $\Delta P$. In such a case,  g~modes of a given angular degree roughly form vertical ridges (analogous to the acoustic modes in the case of solar-like oscillations). On the other hand, rapid rotators show a period spacing that depends approximately linearly on the mode period \citep{VanReeth2016,Ouazzani2017}. Therefore, their period-spacing pattern can be easier identified in a plot of $\Delta P$ as a function of period. For each star, we also complemented such two plots with a periodogram where observed and template periods were overplotted. 

To facilitate the exploration of the parameter space, we developed the interactive tool \texttt{FLOSSY}\footnote{\url{https://github.com/IvS-KULeuven/FLOSSY}}, a Python utility that allows a user to efficiently browse the periodogram of a large number of stars and visualize the period-spacing patterns by displaying the period \'echelle diagram and period-spacing plot at each location in the periodogram. \texttt{FLOSSY} also overplots Eq.\,(\ref{Eq:pattern}) in the aforementioned plots with customized parameters $\{P_0, \Delta P_0, \alpha\}$. The latter can be modified on the fly, along with the number of mode periods to fit. Figure\,\ref{Fig:examples_of_results} shows part of \texttt{FLOSSY}'s output, as well as the best-fit pattern for TIC\,374944608. 

We used \texttt{FLOSSY} to manually select the $\mathbb{\theta^{\textrm{\;init}}}$ for every candidate period-spacing pattern found from the list of mode periods per star. In doing so we considered the parameter space $|P_0-P_0^{\textrm{\;init}}| \le \delta P/2$, $100\ \textrm{s} \le \Delta P_0 \le 4000\ \textrm{s}$ and, $-0.3 \le \alpha \le 0.3$, where $P_0^{\textrm{\;init}} \in \{ P^{\mathrm{\;obs}}_i \}$ and $\delta P$ corresponds to the resolution set in Step 2 of the procedure discussed in Section\,\ref{Sec:freq_analysis}. To ensure that we found a global minimum, we computed $S$ around the best-fit solution in a radius of 400\,s for $P_0$, 40\,s for $\Delta P_0, $ and 0.05 units for $\alpha$. Those values for $S$ are shown in Figure\,\ref{Fig:error_estimate}, which is also an output of  \texttt{FLOSSY}.

\begin{figure}[h]
\centering
\includegraphics[width=\hsize]{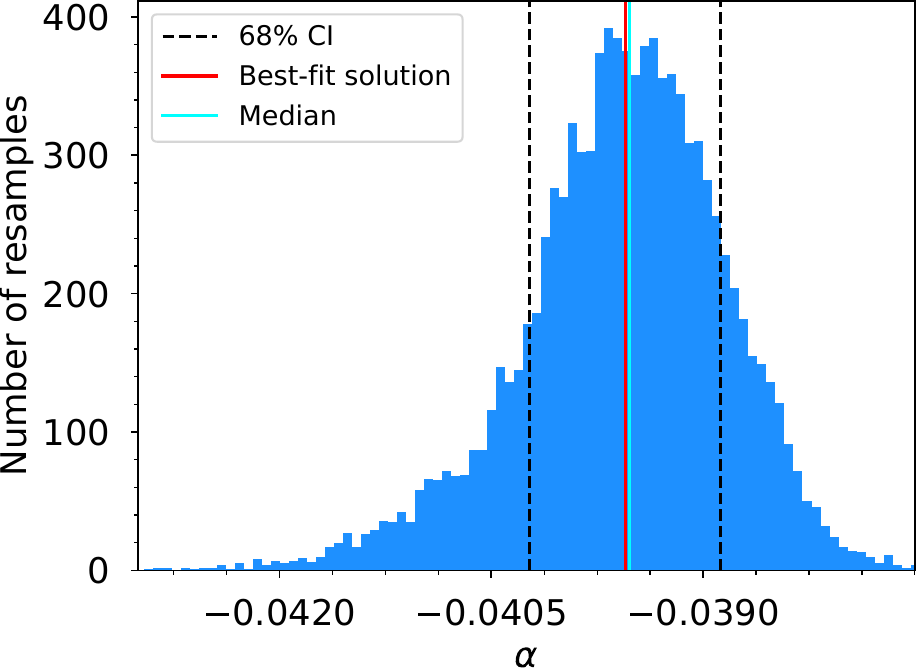}
\caption{Confidence interval (CI) for $\alpha$ following a bootstrap residual procedure for TIC\,374944608. The meaning of the various vertical dashed lines are indicated in the legend.}
\label{Fig:bootstrap}
\end{figure}

To estimate uncertainties for the detected period-spacing pattern, we computed the 68\% confidence interval of the parameters using a bootstrap method with non-parametric residual resampling in the periodogram. We generated 10000 datasets of the same size as the original one. Subsequently, we minimized Eq.\,(\ref{Eq:cost_function}) in each of these datasets using as initial guess the same $\mathbf{\theta}^{\textrm{\;init}}$ as in the original best fit. The confidence intervals were then determined as the 16\% and 84\% quantiles of the bootstrap distribution for the parameters $\mathbf{\theta}$. As an example, Figure\,\ref{Fig:bootstrap} shows the bootstrap distribution of $\alpha$ for the pattern found in TIC\,374944608. 

\section{Catalog of g-mode pulsators in the TESS southern CVZ with identified period-spacing patterns}\label{Sec:PSP-catalog}

\begin{figure}[h!]
\centering
\includegraphics[width=\hsize]{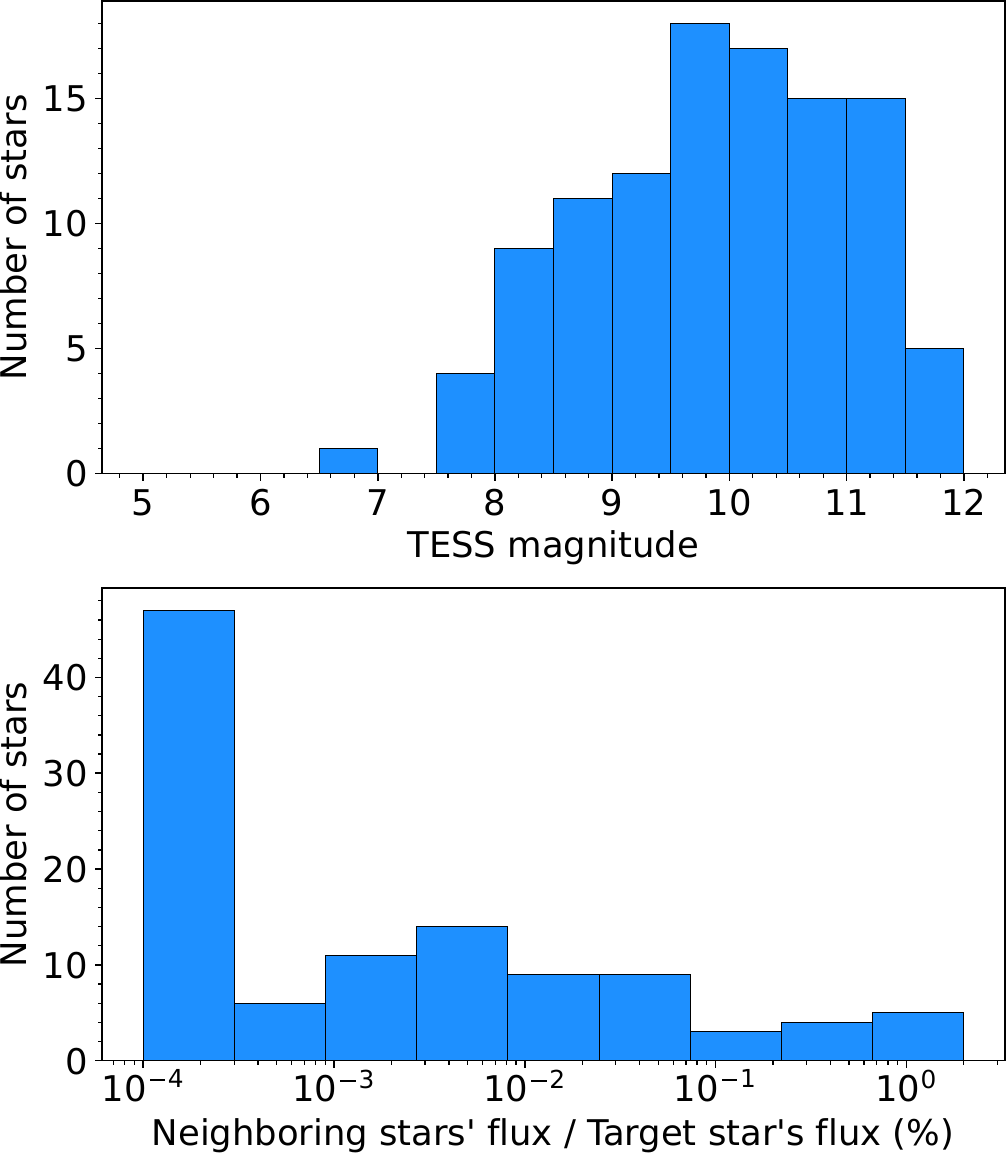}
\caption{Final catalog. {\em Top:}  Apparent TESS magnitude. {\em Bottom:}  Median contamination level in the aperture mask caused by flux of neighboring stars.} 
\label{Fig:hist_Tmag_contamination}
\end{figure}

\begin{figure}[h!]
\centering
\includegraphics[width=\hsize]{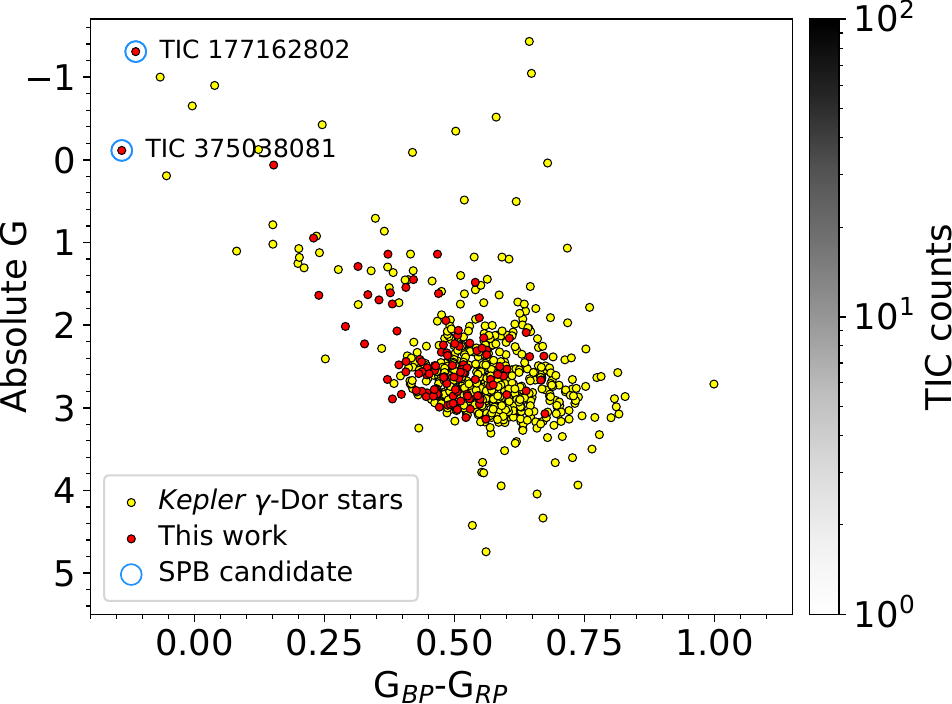}
\caption{{\em Gaia} color-magnitude diagram showing our sample (red) and the 611 $\gamma$-Dor stars with period-spacing patterns found by \citet{GangLi2020} in {\em Kepler} data (yellow). Stars marked with blue circles are SPB candidates. Background stars correspond to the TESS southern CVZ. Magnitudes are not corrected for extinction}
\label{Fig:HRD}
\end{figure}

Our final catalog of g-mode pulsators revealing period-spacing patterns consists of 108 bright dwarfs in the TESS southern CVZ. These stars revealed a total of 140 resolved period-spacing patterns. Each of these patterns are shown in the supplementary material in the same format as Figures\,\ref{Fig:examples_of_results} and \ref{Fig:error_estimate}. Stars in our catalog have apparent TESS magnitudes between 7.5 and 12, with a median of about 10; the star TIC\,350144657 is an exception with an apparent TESS magnitude of about 6.9. The contamination of light curves, due to the flux from neighboring stars, is $<2\%$ thanks to our stringent requirements on the aperture mask described in Section\,\ref{Sec:masks}. Figure\,\ref{Fig:hist_Tmag_contamination} shows the distributions of brightness and contamination in the catalog. Only two stars, TIC\,177162802 and TIC\,375038081, are candidates to be SPB stars, as determined by the color-magnitude selection done in Section\,\ref{Sec:Data_set}, while the other members of the catalog are $\gamma$\,Dor stars. Figure\,\ref{Fig:HRD} shows a {\em Gaia} color-magnitude diagram that compares our catalog to the 611 $\gamma$\,Dor stars with detected period-spacing patterns found by \citet{GangLi2020} in 4-year {\em Kepler} light curves. We find that our catalog stars appear to occur on the hotter end of the {\em Kepler} $\gamma$\,Dor stars because of the uncorrected reddening (T. R. Bedding, priv. comm.).

Out of the 140 period-spacing patterns, 93 have  a downward slope ($\alpha<0$) and 47 have an upward slope ($\alpha>0$). The former are prograde or zonal g~modes while the latter are retrograde g~modes or Rossby modes \citep{VanReeth2016}. The averaged period-spacing value per pattern, $\langle \Delta P \rangle$, is $\sim110$\,s. The shortest pattern contains four measured pulsation periods, more than half of the patterns contain more than 12 and the longest pattern contains 20. In 26\% of the stars we detected two or three patterns. When multiple patterns are detected in a star, this allows for a better constraint of the stellar interior from asteroseismic modeling \citep{Aerts2018}. Furthermore, 29\% of our catalog stars are hybrid pulsators, meaning that they also exhibit p-mode pulsations, therefore, providing us with a means to probe the outer stellar envelope and allowing for a differential study of the star. 

\begin{figure*}[h!]
\centering
\includegraphics[width=\hsize]{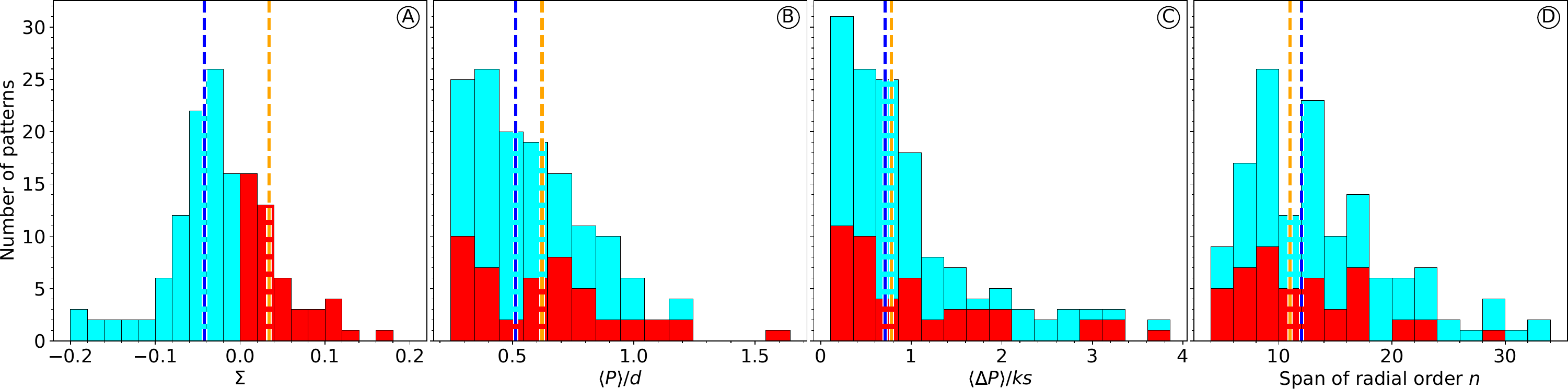}
\caption{Characterization of the best-fit patterns in our sample. The stacked histograms show retrograde modes in red and prograde modes in cyan. The vertical orange and blue lines are the median of the retrograde and prograde distributions, respectively.  \textbf{\Circled{A} :} Slope $\alpha\equiv\mathrm{d}\Delta P/\mathrm{d}P$. \textbf{\Circled{B} :} Mean period $\langle P \rangle$. \textbf{\Circled{C} :} Mean period spacing $\langle \Delta P \rangle$. \textbf{\Circled{D} :} Span of the overtones.}
\label{Fig:hist_patterns_params}
\end{figure*}

\begin{figure}[h!]
\centering
\includegraphics[width=\hsize]{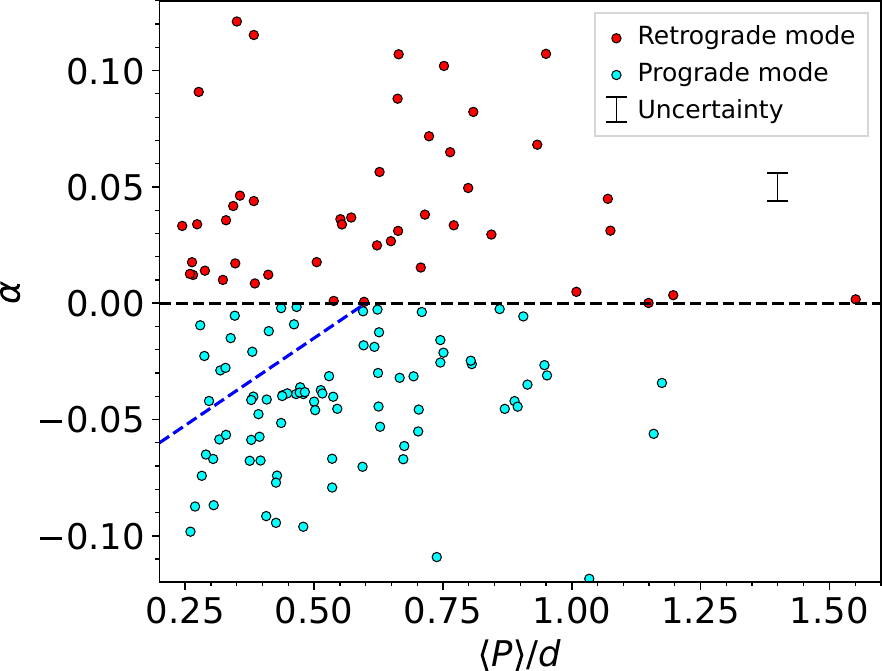}
\caption{$P$-$\alpha$ relation for g-mode pulsators in our catalog. The dashed blue line is the empirical cut from Figure 8 in \citet{GangLi2020} that separates prograde $\ell=1$ g modes (below) from $\ell=2$ (above). The median of the uncertainties has been plotted as a typical uncertainty for $\alpha$. Uncertainties in $\langle P \rangle$ are smaller than the symbol size.}
\label{Fig:P-slope}
\end{figure}

Figure\,\ref{Fig:hist_patterns_params} shows the distributions of the pattern parameters. Typical uncertainties for the pattern parameters are 43 s for $P_0$, 13 s for $\Delta P_0$ and 0.006 for $\alpha$. The pattern slopes fulfill $|\alpha|\le0.1$ for 88\% of the catalog stars, while the tails of the distribution reach $|\alpha|\sim0.2$. Using the empirical relations in \citet{GangLi2020}, we can estimate the near-core rotation to be $<1.68\,\textrm{d}^{-1}$ for 86\% of the stars in our sample, with a few stars reaching up to about 2.86\,$\textrm{d}^{-1}$. We also made use of the Figure 8 in \citet{GangLi2020} to define the empirical cut delineating regimes of dipole and quadrupole modes in Figure\,\ref{Fig:P-slope} (dashed blue line). This suggests that 13 of our prograde patterns have $\ell=2,$ while the rest of them have $\ell=1$. We noted that because pulsations in $\gamma$\,Dor stars are sensitive to metallicity, the empirical estimates above drawn using \citet{GangLi2020} remain to be confirmed. Since the nominal {\em Kepler} field-of-view was in the northern hemisphere, we cannot directly cross-validate our TESS southern CVZ results with those from {\it Kepler}. 

We found a positive correlation for prograde modes with $l=1$ between the parameters $\langle P \rangle$ and $\alpha$ with a Pearson correlation coefficient of 0.67 and a $p$-value of $10^{-9}$, while prograde modes with $l=2$ show the same Pearson correlation coefficient but with a $p$-value of 0.016. Since $\langle P \rangle$ is a proxy for the evolutionary stage, these correlations reveal that the near-core rotation rate of the stars slows down as they evolve, implying that an efficient angular momentum transport mechanism must be at work as already found in the literature \citep[cf.\,][for a review]{Aerts2019}. We did not search for a correlation in retrograde modes because their parameter $\alpha$ is less sensitive to the star’s local rotation rate. For those,  a more precise analysis involving the traditional approximation of rotation \citep[TAR;][]{Eckart1960} is necessary and will be addressed in a future paper. Furthermore, the range in overtones in the patterns is a proxy for the radial order $n$ of the g~modes. The exact value of $n$, $\ell,$ and $m$ can only be identified from asteroseismic modeling, for example based on the TAR as applied in \citet{VanReeth2016}. The mode identification and asteroseismic modeling based on the TAR will be addressed in a future paper dedicated to the ensemble of stars in our catalog of g-mode pulsators, relying on the pattern properties deduced in this work.

Besides quasi-linear period-spacing patterns like the one shown in panel \textbf{\Circled{C}} of Figure\,\ref{Fig:examples_of_results}, where the zigzag feature is caused by missing periods in the pattern (white circles), our catalog contains tens of patterns with zigzag signatures that are not related to missing modes. These patterns are presented in the supplementary material, where Figures \ref{Fig:examples_of_results} and \ref{Fig:error_estimate} are reproduced for each period-spacing pattern in our catalog. Such signatures have also been  observed in period-spacing patterns of SPB stars, where they are interpreted as the result of strong envelope mixing deep inside the star. These signatures were recently used by \citet{Pedersen2021} to constrain the internal mixing profile in SPB stars observed by {\em Kepler}. The levels of envelope mixing found thus far in $\gamma\,$Dor stars are far lower than those of SPB stars \citep[cf.\,Table\,1 in ][]{Aerts2021}. Our catalog presents an opportunity to further assess and refine these recent conclusions in the literature from our TESS southern CVZ catalog of g-mode dwarf pulsators. We also noted that many of the detected period-spacing patterns in our catalog show residuals with a sinusoidal-like modulation after the subtraction of the linear fit (red line in panel \textbf{\Circled{C}} of Figure\,\ref{Fig:examples_of_results}). This type of periodic residuals is very similar to the one found originally for the slowly rotating CoRoT SPB HD\,50230 by \citet{Degroote2010} and allows for stringent constraints on the core overshooting and envelope mixing. 

Finally, we compared our catalog to the spectroscopic parameters published in the GALAH Data Release (DR) 3 paper by \citet{Buder2021}. This DR3 has 38 stars in common with our sample. For these 38 stars, we found the effective temperature, surface gravity, and surface velocity deduced from line-profile broadening to be consistent with such properties of $\gamma$\,Dor stars \citep[e.g.,][]{VanReeth2015b,GangLi2020}, that is, early F- to late A-type main-sequence stars. This agreement in stellar parameters supports the selection methods used throughout the current paper. The corresponding distributions can be found in Appendix\,A, where we also discuss the correlation found between the surface velocity estimate from spectral line broadening and the average pulsation period $\langle P \rangle$. As expected for moderate- to fast-rotating pulsators, the larger surface velocity is accompanied by a shorter average pulsation period.

\section{Summary}

In this work, we present a new data analysis pipeline to create light curves from TESS full frame images (FFI), with an emphasis on the search for g-mode frequencies in intermediate- to high-mass stars. We present guidelines for extracting light curves from unprocessed TESS images, including the selection of aperture and background masks, the identification of time stamps affected by systematics in sectors 1-13 of the TESS southern CVZ, and a modified PCA to detrend the light curves. A color-magnitude criterion was presented as a method to identify main-sequence A/F- and O/B-type star candidates. We also introduced \texttt{FLOSSY}, an open source utility for inspecting periodograms of g-mode pulsators to facilitate searches for period-spacing patterns.

Based on the light curves extracted with our pipeline, we composed the first catalog of g-mode period-spacing patterns detected in TESS space photometry of dwarfs having colors representative of spectral types F to O. Our catalog contains 140 g-mode period-spacing patterns observed in 106 $\gamma$\,Dor stars and 2 SPB stars. The patterns were manually reviewed and contain g-mode frequencies having amplitudes of S/N>4. In a future work, we will use the detected patterns to derive the internal rotation frequency near the convective core of the stars, as well as the buoyancy travel time across the stars (known as $\Pi_0$). These two key quantities constitute important observables that are useful for performing asteroseismic modeling of intermediate-mass stars \citep[e.g.,][]{Szewczuk2018,Mombarg2021,Pedersen2021}.  
The nominal frequency resolution of modes in the detected patterns amounts to $0.003\textrm{d}^{-1}$, following the 352\,d long TESS southern CVZ light curves. This frequency resolution can be improved by a factor of 3 when the extended TESS Cycle 3 data will be included in the analysis. This will also lower the noise level in the Fourier domain and offers the future potential to detect more modes per star, as well as more stars with g-mode patterns. 
Our catalog and global properties of the detected patterns in it are publicly available through
CDS. The patterns themselves, are shown in the supplementary material. This catalog constitutes a base for future ensemble asteroseismic modeling of TESS g-mode pulsators following methodologies as in \citet{Mombarg2021} or \citet{Pedersen2021}. In this way, we will be able  to constrain the internal physics of more rotating dwarfs with a convective core using the new available TESS data, in addition to the asteroseismic modeling achieved so far for a legacy sample of g-mode pulsators from the {\it Kepler\/} mission \citep{Gebruers2021}. This will increase the number of dwarfs with such modeling and will lead to a better understanding of the transport processes and their relationship to the internal rotation profile of these stars.

\begin{acknowledgements}

The research leading to these results has received funding from the European Research Council (ERC) under the European Union’s Horizon 2020 research and innovation programme (grant agreement no. 670519: MAMSIE) and from 
the KU\,Leuven Research Council (grant C16/18/005: PARADISE). TVR gratefully acknowledges support from the Research Foundation Flanders under grant agreement nr. 12ZB620N. This research has made use of the SIMBAD database, operated at CDS, Strasbourg, France. This work made use of the Third Data Release of the GALAH Survey (Buder et al. 2021). The GALAH Survey is based on data acquired through the Australian Astronomical Observatory, under programs: A/2013B/13 (The GALAH pilot survey); A/2014A/25, A/2015A/19, A2017A/18 (The GALAH survey phase 1); A2018A/18 (Open clusters with HERMES); A2019A/1 (Hierarchical star formation in Ori OB1); A2019A/15 (The GALAH survey phase 2); A/2015B/19, A/2016A/22, A/2016B/10, A/2017B/16, A/2018B/15 (The HERMES-TESS program); and A/2015A/3, A/2015B/1, A/2015B/19, A/2016A/22, A/2016B/12, A/2017A/14 (The HERMES K2-follow-up program). We acknowledge the traditional owners of the land on which the AAT stands, the Gamilaraay people, and pay our respects to elders past and present. This paper includes data that has been provided by AAO Data Central (datacentral.aao.gov.au). This work has made use of data from the European Space Agency (ESA) mission {\it Gaia} (\url{https://www.cosmos.esa.int/gaia}), processed by the {\it Gaia}
Data Processing and Analysis Consortium (DPAC, \url{https://www.cosmos.esa.int/web/gaia/dpac/consortium}). Funding for the DPAC has been provided by national institutions, in particular the institutions
participating in the {\it Gaia} Multilateral Agreement. In addition to the software cited in the main body of the paper we have also made use of \texttt{Lightkurve}, a Python package for {\em Kepler} and TESS data analysis \citep{lightkurve}, \texttt{Astropy},\footnote{http://www.astropy.org} a community-developed core Python package for Astronomy \citep{astropy:2013, astropy:2018}, \texttt{Matplotlib} \citep{Matplotlib}, \texttt{NumPy} \citep{NumPy}, \texttt{SciPy} \citep{SciPy}, \texttt{pandas} \citep{Pandas}. Finally, we would like to thank the anonymous referee for useful comments and suggestions.

\end{acknowledgements}

  \bibliographystyle{aa} 
  \bibliography{aanda} 

\begin{appendix}

\section{Extended tables and figures} 

Table\,\ref{Tab:excluded_intervals} lists the removed time intervals before the detrending of the light curves described in Section \ref{subsec:lc}. The best-fit values of the parameters defined in Section \ref{Sec:PSP-seach} that characterize the 140 period-spacing patterns in our catalog are publicly available through CDS. The supplementary material available online contains a version of Figures\,\ref{Fig:examples_of_results} and \ref{Fig:error_estimate} that show the periodogram and the best-fit period-spacing pattern for all stars in our catalog. Whenever periods are plotted with gray circles in those figures, those periods are ignored in the fit.

\begin{figure}[h!]
\includegraphics[width=\hsize,clip]{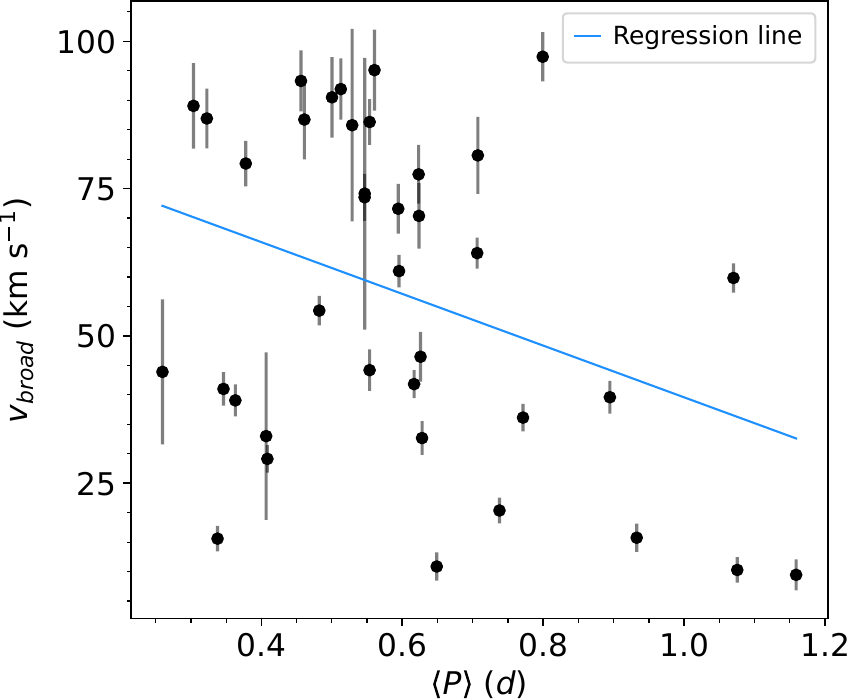}
\caption{Correlation between the surface velocity deduced from spectral line broadening as reported by GALAH DR3 and our pattern parameter $\langle P \rangle$. If a star has multiple patterns, then the pattern with the highest-amplitude pulsations was used. The plot contains the 38 stars in common in GALAH DR3 and our catalog, where we only considered stars with spectra flagged as reliable in GALAH DR3. Uncertainties in $\langle P \rangle$ are smaller than the symbol size, while those of the velocities are underestimates for reasons explained in the text.}
\label{Fig:GALAH_correlations}
\end{figure}

\begin{figure*}
\resizebox{\hsize}{!}
{\includegraphics[width=\hsize,clip]{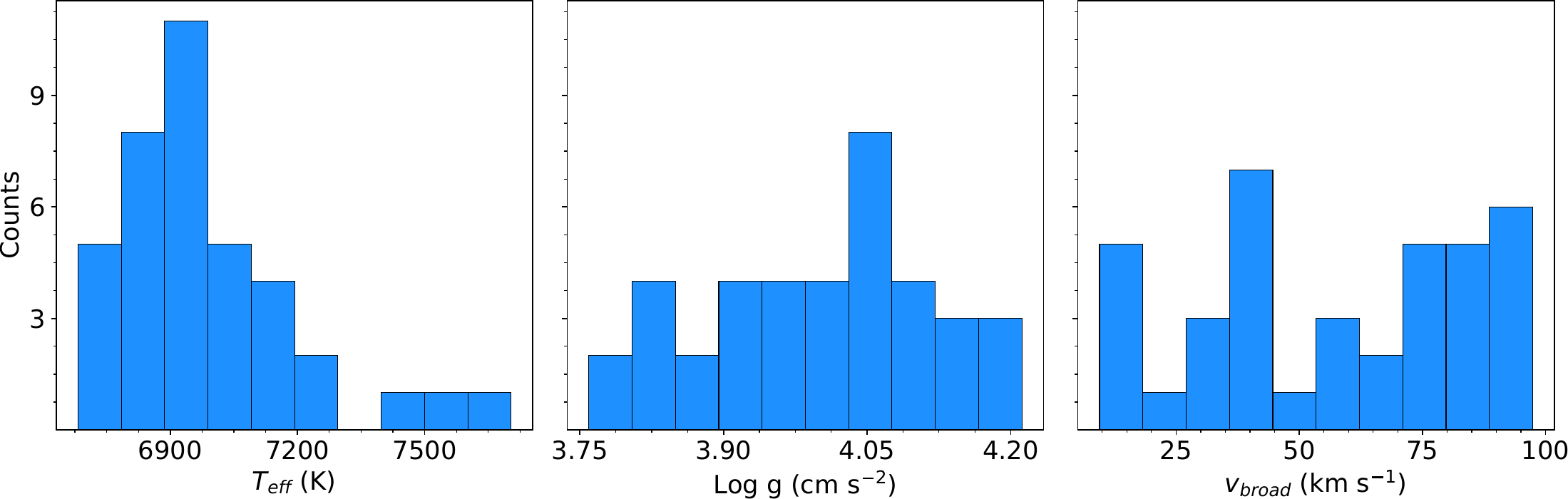}}
\caption{Histograms for the spectroscopic stellar parameters of the 38 stars in common with our catalog having spectra flagged as reliable in the GALAH DR3 catalog by \citet{Buder2021}.}
\label{Fig:GALAH_hist}
\end{figure*}

Figure\,\ref{Fig:GALAH_correlations} shows a negative correlation with a Pearson correlation coefficient of -0.34 and a p-value of 0.04 between the  mean pulsation period of our catalog stars and their surface velocity estimated from spectral line broadening as reported in GALAH DR3 \citep{Buder2021}. In other words, larger surface velocities are measured for stars with shorter $\langle P \rangle$. While the measurement of the surface velocity from spectral line broadening is complicated and can never be precise in the presence of time-dependent gravity-mode velocity broadening \citep{Aerts2014}, this result is consistent with the expected effect of faster rotation on the periods of the pulsation modes \citep{Bouabid2013}. There is significant scatter in the figure, as expected for a sample of stars seen at random inclination angles, given the small overlap of only 38 stars between our catalog and GALAH DR3. Moreover, the uncertainties reported by GALAH DR3 are underestimates given that the spectral line broadening changes considerably throughout the pulsation cycles as shown by \citet{Aerts2014}, while that time dependence has been ignored in the reported velocity estimates based on snapshot spectra. This  precludes us to connect in-depth conclusions based on this correlation in Fig.\,\ref{Fig:GALAH_correlations}. Figure\,\ref{Fig:GALAH_hist} shows the distributions of mass, surface gravity, and effective temperature for our catalog stars available in GALAH DR3. The ranges of these three parameters are in agreement with those of $\gamma$\,Dor stars \citep{VanReeth2015b,GangLi2020}.

\begin{table}[h]
    \centering
    \caption[]{Intervals excluded from the light curves.}
    {\renewcommand{\arraystretch}{1.3} 
    \begin{tabular}{cc}
    \hline \hline
    TESS sector         & Excluded interval \\ 
    & (BJD-2457000) \\
    \hline 
    \multirow{2}{*}{1}  & (1334.8, 1335.1)             \\
                        & (1347.0, 1349.5)             \\ \hline
    \multirow{7}{*}{2}  & (1356.2, 1356.5)             \\
                        & (1361.0, 1361.3)             \\
                        & (1363.5, 1363.8)             \\
                        & (1365.9, 1366.2)             \\
                        & (1373.8, 1374.1)             \\
                        & (1375.8, 1376.0),            \\
                        & (1377.9, 1378.7)             \\ \hline
    \multirow{10}{*}{3} & (1380.0, 1385.0)             \\
                        & (1387.6, 1387.9)             \\
                        & (1390.1, 1390.4)             \\
                        & (1392.6, 1392.9)             \\
                        & (1395.1, 1395.4)             \\
                        & (1398.6, 1398.9)             \\
                        & (1400.6, 1400.9)             \\
                        & (1402.6, 1402.9)             \\
                        & (1404.6, 1404.9)             \\
                        & (1406.1, 1406.4)             \\ \hline
    4                   & (1420.0, 1427.0)             \\ \hline
    5                   & (1463.0, 1465.0)             \\ \hline
    6                   & (1476.0, 1479.0)             \\ \hline
    7                   & (1502.5, 1506.0)             \\ \hline
    8                   & (1529.5, 1533.0)             \\ \hline
    \end{tabular}
    }
    \tablefoot{Intervals excluded from the light curves to remove systematic flux variability due to, e.g., scattered light, telescope jittering, and loss of fine pointing.}
    \label{Tab:excluded_intervals}
\end{table}

\end{appendix}

\end{document}